\pdfoutput=1

\documentclass[twocolumn]{aastex631}

\usepackage{amsmath}
\usepackage{amssymb}
\usepackage{hyperref}
\usepackage{color}
\usepackage{lmodern}
\usepackage{graphicx}
\usepackage{xspace}

\newcommand{\code}[1]{\texttt{#1}}

\newcommand{\AthenaPP}{\code{Athena++}\xspace}
\newcommand{\AthenaK}{\code{AthenaK}\xspace}
\newcommand{\GRAthena}{\code{GR-Athena++}\xspace}
\newcommand{\HARM}{\code{HARM}\xspace}
\newcommand{\WhiskyTHC}{\code{WhiskyTHC}\xspace}
\newcommand{\EinsteinToolkit}{\code{EinsteinToolkit}\xspace}

\newcommand{\Kokkos}{\code{Kokkos}\xspace}
\newcommand{\RePrimAnd}{\code{RePrimAnd}\xspace}
\newcommand{\PrimitiveSolver}{\code{PrimitiveSolver}\xspace}
\newcommand{\GRaMX}{\code{GRaM-X}\xspace}
\newcommand{\AsterX}{\code{AsterX}\xspace}

\newcommand{\LORENE}{\code{LORENE}\xspace}
\newcommand{\SpECTRE}{\code{SpECTRE}\xspace}
\newcommand{\DendroGR}{\code{Dendro-GR}\xspace}
\newcommand{\ECHO}{\code{ECHO}\xspace}
\newcommand{\BHAC}{\code{BHAC}\xspace}
\newcommand{\IllinoisGRMHD}{\code{IllinoisGRMHD}\xspace}
\newcommand{\CosmosPP}{\code{Cosmos++}\xspace}
\newcommand{\HARMNoble}{\code{HARM-Noble}\xspace}
\newcommand{\iharm}{\code{iharm3D}\xspace}
\newcommand{\KORAL}{\code{KORAL}\xspace}
\newcommand{\Parthenon}{\code{Parthenon}\xspace}

\newcommand{\Msun}{\textrm{M}_\odot}
\newcommand{\Hz}{\textrm{Hz}}
\newcommand{\atm}{\textrm{atm}}
\newcommand{\thr}{\textrm{thr}}

\newcommand{\tmax}{\textrm{max}}

\newcommand{\intrad}[1]{\int_0^{2\pi}\int_0^\pi{#1}d\theta d\phi}

\begin{document}

\title{Performance-Portable Binary Neutron Star Mergers with AthenaK}

\author[0000-0001-5705-1712]{Jacob \surname{Fields}}
\email{jmf6719@psu.edu}
\affiliation{Department of Physics, The Pennsylvania State 
University, University Park, PA 16802}
\affiliation{Institute for Gravitation \& the Cosmos, The
Pennsylvania State University, University Park, PA 16802}

\author[0000-0001-9027-4184]{Hengrui \surname{Zhu}}
\affiliation{Department of Physics, Princeton University, Jadwin Hall, Washington Road, New Jersey, 08544, USA}
\affiliation{Princeton Gravity Initiative, Princeton University, Princeton, New Jersey, 08544, USA}

\author[0000-0001-6982-1008]{David \surname{Radice}}\thanks{Alfred P.~Sloan Fellow}
\affiliation{Department of Physics, The Pennsylvania State 
University, University Park, PA 16802}
\affiliation{Department of Astronomy \& Astrophysics, The Pennsylvania State 
University, University Park, PA 16802}
\affiliation{Institute for Gravitation \& the Cosmos, The
Pennsylvania State University, University Park, PA 16802}

\author[0000-0001-5603-1832]{James M.~\surname{Stone}}
\affiliation{School of Natural Sciences, Institute for Advanced Study, 1 Einstein Drive, Princeton, NJ 08540, USA}

\author[0000-0003-2244-3462]{William \surname{Cook}}
\affiliation{Theoretisch-Physikalisches Institut, Friedrich-Schiller-Universit{\"a}t Jena, 07743, Jena, Germany}

\author[0000-0002-2334-0935]{Sebastiano \surname{Bernuzzi}}
\affiliation{Theoretisch-Physikalisches Institut, Friedrich-Schiller-Universit{\"a}t Jena, 07743, Jena, Germany}

\author[0000-0001-6091-2827]{Boris \surname{Daszuta}}
\affiliation{Theoretisch-Physikalisches Institut, Friedrich-Schiller-Universit{\"a}t Jena, 07743, Jena, Germany}

\date{\today}

\begin{abstract}
  We introduce an extension to the \AthenaK code for general-relativistic
  magnetohydrodynamics (GRMHD) in dynamical spacetimes using a 3+1 conservative Eulerian
  formulation. Like the fixed-spacetime GRMHD solver, we use standard finite-volume
  methods to evolve the fluid and a constrained transport scheme to preserve the
  divergence-free constraint for the magnetic field. We also utilize a first-order flux
  correction (FOFC) scheme to reduce the need for an artificial atmosphere and optionally
  enforce a maximum principle to improve robustness. We demonstrate the accuracy of
  \AthenaK using a set of standard tests in flat and curved spacetimes. Using a SANE
  accretion disk around a Kerr black hole, we compare the new solver to the existing
  solver for stationary spacetimes using the so-called ``HARM-like'' formulation. We find
  that both formulations converge to similar results. We also include the first published
  binary neutron star (BNS) mergers performed on graphical processing units (GPUs). Thanks
  to the FOFC scheme, our BNS mergers maintain a relative error of $\mathcal{O}(10^{-11})$
  or better in baryon mass conservation up to collapse. Finally, we perform scaling tests
  of \AthenaK on OLCF Frontier, where we show excellent weak scaling of $\geq 80\%$
  efficiency up to 32768 GPUs and $74\%$ up to 65536 GPUs for a GRMHD problem in dynamical
  spacetimes with six levels of mesh refinement. \AthenaK achieves an order-of-magnitude
  speedup using GPUs compared to CPUs, demonstrating that it is suitable for performing
  numerical relativity problems on modern exascale resources.
\end{abstract}


\section{Introduction}
\label{sec:intro}
Over the last few decades, numerical relativity (NR) has matured into a robust tool for
making accurate quantitative predictions for astrophysical phenomena in the strong-field
regime of general relativity. Notable successes include the first detection of
gravitational waves (GW150914) and the first observed binary neutron star merger
(GW170817), both of which were in excellent agreement with models informed by numerical
relativity \citep{LIGOScientific:2016aoc,LIGOScientific:2017vwq}. Observational data from
GW170817 in particular has further fed back into NR, which has been able to use targeted
simulations to place constraints on the neutron star equation of state (EOS) and other
progenitor characteristics
\citep{Shibata:2017xdx,Ruiz:2017due,Radice:2017lry,Hinderer:2018pei}.

Nevertheless, current NR codes face a number of challenges going forward. Improvements in
existing detectors and proposed next-generation detectors promise unprecedented accuracy
in observing gravitational-wave events \citep{LIGO.7.16,Maggiore.3.2020}. However, such
improvements will require more accurate waveform models, particularly for BNS systems,
which are currently dominated by systematic errors at high signal-to-noise ratios
\citep{Purrer:2019jcp,Gamba:2020wgg}.

We also expect future detectors to be sensitive to the BNS post-merger phase. NR data in
this regime exhibits a number of issues, with features such as the merger time, peak
frequency, and survival time depending heavily on numerical schemes, choice and
implementation of microphysics, and resolution \citep{Espino:2022mtb,Zappa:2022rpd}.

These demands all point to the need for NR simulations at higher resolutions with
more accurate numerical methods and more realistic physics. Some recent efforts include
the use of high-order methods (including finite-difference, discontinuous Galerkin
methods, and pseudospectral methods where appropriate)
\citep{PhysRevD.93.063006,PhysRevD.94.084004,Kidder:2016hev,Most:2019kfe,Tichy:2022hpa,
Deppe:2023qxa},
more accurate Riemann solvers and treatments of the divergence-free condition
\citep{Kiuchi:2022ubj}, and more realistic neutrino physics
\citep{Foucart:2021mcb,Radice:2021jtw,Izquierdo:2022eaz}.

Just as important is implementing new computational algorithms which allow for better
parallel scaling and more efficient use of resources. For example, both \SpECTRE and
\GRAthena, though very different in implementation, use task-based parallelism to overlap
communication and computation \citep{Kidder:2016hev,Stone.7.20,Daszuta:2021ecf,
Cook:2023bag}.
One more notable example is block-based adaptive mesh refinement (AMR). By dividing the
mesh into a series of blocks (typically stored in an octree data structure), mesh
refinement is achieved by replacing a single block with eight smaller blocks. Though the
need to refine entire blocks (often with a 2:1 size constraint) means that more
traditional patch-based AMR approaches are in principle more flexible, block AMR avoids
redundant computation and is much simpler to parallelize efficiently. Consequently,
block-based codes, such as \DendroGR \citep{Fernando:2018mov,Fernando:2022php} and
\GRAthena \citep{Stone.7.20,Daszuta:2021ecf,Cook:2023bag,Rashti:2023wfe}, typically
exhibit excellent scaling properties up to $\mathcal{O}(10^4)$ cores or more, and the
simpler remeshing and load-balancing operations mean that in practice their grid
refinement compares favorably with the most adaptive patch-based methods (e.g.,
\citet{Clough:2015sqa,Radia:2021smk}).

Another approach is to use graphical processing units (GPUs). Due to their higher energy
efficiency compared to CPUs, many modern supercomputers derive a major portion of their
computational power from GPUs. Recently, \DendroGR showed the ability to accelerate binary
black hole calculations with GPUs \citep{Fernando.11.22}, and the \GRaMX and \AsterX codes
demonstrated single neutron star evolutions in the \EinsteinToolkit with
GPUs \citep{Shankar.10.22,Kalinani:2024rbk}.

One challenge for GPU development is that there exist multiple GPU vendors which each
preferentially support different programming models, such as CUDA and HIP. Furthermore,
many new machines still contain significant CPU resources, and many tasks such as
debugging and simple testing may be easier to do on a personal laptop or workstation,
which may not necessarily have a GPU. This leads to a desire for \textit{performance
portability}, or the ability not only to run a piece of software on a wide variety of
machine architectures and software stacks but to do so \textit{well}. However, manually
rewriting different parts of an application for each supported architecture is
time-consuming and error-prone.

A more convenient route to performance portability is to use pre-existing libraries.
Though several tools exist for performance portability, the one that we highlight here is
the \Kokkos C++ framework \citep{Kokkos}. \Kokkos is an abstraction layer which hides
platform-specific details such as optimal memory layout and launching GPU or OpenMP
kernels, enabling developers to write high-performance scientific code once and compile on
a variety of architectures with various programmming models.



\AthenaK\footnote{see \url{https://github.com/IAS-Astrophysics/athenak}} is a complete
rewrite of the \AthenaPP code using \Kokkos. This paper is the third in a series
describing its features and capabilities, with \citet{Stone.6.24} focusing on its
architecture and applications to astrophysical fluids, \citet{Zhu.6.24} discussing the
dynamical spacetime solver and infrastructure for binary black hole (BBH) problems, and
this paper discussing an extension which couples GRMHD to dynamical spacetimes. In
Section~\ref{sec:mhd}, we briefly discuss the GRMHD equations.
Section~\ref{sec:nummethods} provides an overview of our numerical methods, with
particular attention given to our atmosphere treatment and the implementation of a
first-order flux correction (FOFC) scheme. We demonstrate the code's accuracy and
robustness in Section~\ref{sec:tests}, including a convergence test with the first
published BNS mergers performed on GPUs. Lastly, Section~\ref{sec:scaling} describes the
results of our performance and scaling tests, demonstrating \AthenaK's suitability for
exascale machines. Lastly, we summarize our work in Section~\ref{sec:conclusion} and
highlight some potential new NR applications for \AthenaK.

Throughout the paper, we follow the standard indexing convention: Latin
indices refer to spatial indices, e.g., $i,j,k... = \{1, 2, 3\}$, and Greek indices
are spacetime indices, e.g., $\mu,\nu,... = \{0, 1, 2, 3\}$. We also adopt geometric units
such that $G=c=k_\mathrm{B}=1$.

\section{3+1 Relativistic Magnetohydrodynamics}
\label{sec:mhd}
The ideal GRMHD equations are described by the following systems of equations:
\begin{subequations}
\label{eq:cov_mhd}
\begin{align}
  \label{eq:barycons}
  \nabla_\mu\left(\rho u^\mu\right) &= 0, \\
  \label{eq:stencons}
  \nabla_\mu T^{\mu\nu} &= 0, \\
  \label{eq:gaussfaraday}
  \nabla_\mu \left( \ast F\right)^{\mu\nu} &= 0,
\end{align}
\end{subequations}
where the first represents baryon conservation; the second, conservation of stress-energy;
and the third, the Gauss-Faraday law. The stress-energy tensor for this system is
\begin{equation}
  \label{eq:Tmunu}
  T^{\mu\nu} = \left(\rho h + b^2\right)u^\mu u^\nu + \left(P + \frac{b^2}{2}\right)
               g^{\mu\nu} - b^\mu b^\nu,
\end{equation}
for a fluid with rest-mass density $\rho$, four-velocity $u^\mu$, pressure $P$, total
fluid energy density $e$, and magnetic field (measured in the comoving frame) $b^\mu$ with
a spacetime metric $g_{\mu\nu}$. We further define the total specific enthalpy 
$h \equiv \left(e + P\right)/\rho$, and we write the dual of the electromagnetic tensor as
\begin{equation}
  \label{eq:Fmunu}
  \left(\ast F\right)^{\mu\nu} = b^\mu u^\nu - b^\nu u^\mu.
\end{equation}

Like its predecessor, \AthenaPP, the standard GRMHD solver in \AthenaK uses a formulation
similar to \citet{Gammie:2003rj} (hereafter referred to as ``\HARM-like''). By
lowering the free index in Eq.~\ref{eq:stencons}, the GRMHD equations take the form
\begin{subequations}
\label{eq:harm_eqs}
\begin{align}
  \label{eq:harm_bary}
  \partial_t\left(\sqrt{-g}\rho u^0\right) + 
    \partial_j\left(\sqrt{-g}\rho u^j\right) &= 0, \\
  \label{eq:harm_sten}
  \partial_t\left(\sqrt{-g}T^0\!_\mu\right) +
    \partial_j\left(\sqrt{-g}T^j\!_\mu\right) &= 
    \frac{1}{2}\sqrt{-g}\left(\partial_\mu g_{\alpha\beta}\right)T^{\alpha\beta}, \\
  \label{eq:harm_gf}
  \partial_t\left(\sqrt{-g}\hat{B}^i\right) + 
    \partial_j\left(\sqrt{-g}\left(\ast F\right)^{ij}\right) &= 0,
\end{align}
\end{subequations}
where $\hat{B}^i$ is the magnetic field measured in the coordinate frame (i.e.,
$\hat{B}^i = \left(\ast F\right)^{i0}$), and $g=\det \mathbf{g}$.
The source terms in Eq.~\ref{eq:harm_sten} vanish for ignorable coordinates, which leads
to better conservation of energy in fixed spacetimes. However, this is no longer true in
dynamical spacetimes, and the structure of this source term is inconvenient.

To formulate a valid Cauchy problem, the Einstein equations are rewritten via a 3+1
decomposition. In the ADM formulation, the line element
$ds^2 \equiv g_{\mu\nu}dx^\mu dx^\nu$ is rewritten as
\begin{equation}
\label{eq:adm_metric}
  ds^2 = -\alpha^2 dt^2 
    + \gamma_{ij}\left(dx^i + \beta^i dt\right)\left(dx^j + \beta^j dt\right),
\end{equation}
where $\alpha$ is the lapse, $\beta^i$ is the shift, and $\gamma_{ij}$ is the metric of a
spatial slice $\Sigma_t$, whose embedding in spacetime is described by the extrinsic
curvature $K_{ij}$. While the evolution equations for $\gamma_{ij}$ and $K_{ij}$ are
derived as part of the decomposition, both $\alpha$ and $\beta^i$ are gauge variables with
some amount of freedom in their specification. Consequently, the
$\partial_t g_{\alpha\beta}$ source term for $T^0\!_0$ in Equation~\ref{eq:harm_sten}
contains $\partial_t \alpha$ and $\partial_t \beta_i$, which cannot be generally
eliminated without prior knowledge of the gauge condition.

For problems in dynamical spacetimes, the so-called ``Valencia'' formulation
\citep{1997ApJ...476..221B,Anton:2005gi} is a more natural choice for GRMHD. We will
provide a brief description of this formalism here, but we advise the reader to consult
the literature for a more complete discussion (e.g.,
\citet{Giacomazzo:2007ti,Cook:2023bag}).

We define a set of conserved variables $\mathbf{U} = \left\{D, S_i, \tau\right\}$:
\begin{subequations}
\label{eq:conserved}
\begin{align}
  \label{eq:dens}
  D &= \rho W, \\
  \label{eq:mom}
  S_i &= \left(\rho h W^2 + B^2\right)v_i - \left(B^k v_k\right)B_i, \\
  \label{eq:tau}
  \tau &= \rho h W^2 + B^2 - P - 
    \frac{1}{2}\left[\left(B^k v_k\right)^2 + \frac{B^2}{W^2}\right] - D,
\end{align}
\end{subequations}
where $v^i$ is the three-velocity in the Eulerian frame, $W = 1/\sqrt{1 - v^2}$ is the
Lorentz factor, and $B^i = n_\mu\left(\ast F\right)^{\mu i}$ is the magnetic field in
the Eulerian frame. The relationship between $b^\mu$ and $B^i$ is summarized as
\begin{subequations}
\label{eq:Btob}
\begin{align}
  b^0 &= \frac{W B^i v_i}{\alpha}, \\
  b^i &= \frac{B^i + \alpha b^0 u^i}{W}.
\end{align}
\end{subequations}

For notational ease, we also use the tilde $\sim$ to designate a \textit{densitized}
variable $\tilde{A} = \sqrt{\gamma} A$, and we define
$\hat{v}^i = u^i/W = v^i - \beta^i/\alpha$ and the total pressure
$P_\textrm{tot} = P + b^2/2$. The GRMHD equations then take the form
\begin{subequations}
\label{eq:val_eqs}
\begin{align}
  \label{eq:val_D}
  \partial_t\tilde{D} + \partial_j\left(\alpha\tilde{D}\hat{v}^j\right) 
    &= 0, \\
  \label{eq:val_S}
  \partial_t\tilde{S_i} + \partial_j\left(\alpha\tilde{S}_i \hat{v}^j -
    \alpha b_i \frac{\tilde{B}^j}{W} + 
    \alpha\tilde{P}_\textrm{tot}\delta^j\!_i\right) 
    &= G^S_i\left(\mathbf{U}\right), \\
  \label{eq:val_tau}
  \partial_t\tilde{\tau} + \partial_j\left(\alpha\tilde{\tau}\hat{v}^j -
    \alpha^2 b^0 \frac{\tilde{B}^j}{W} + 
    \alpha\tilde{P}_\textrm{tot}v^j\right)
    &= G^\tau\left(\mathbf{U}\right), \\
  \label{eq:val_ind}
  \partial_t\tilde{B}^i + \partial_j\left(\alpha\left[\tilde{B}^i\hat{v}^j
    - \tilde{B}^j\hat{v}^i\right]\right) &= 0,
\end{align}
\end{subequations}
with the source terms
\begin{subequations}
\label{eq:val_srcs}
\begin{align}
  G^S_i\left(\mathbf{U}\right) &= \frac{1}{2}\alpha\tilde{S}^{jk}\partial_i
    \gamma_{jk} + \tilde{S}_j \partial_i\beta^j - \tilde{E}\partial_i\alpha,\\
  G^\tau\left(\mathbf{U}\right) &= \alpha K_{ij}\tilde{S}^{ij} -
    \tilde{S}^i\partial_i\alpha,
\end{align}
\end{subequations}
where $E \equiv \tau + D$ is the total energy density in the Eulerian frame, and
$S_{ij} \equiv T_{ij}$ is the stress tensor. Since $B^i = \alpha \hat{B}^i$, it follows
that $\tilde{B}^i = \sqrt{-g}\hat{B}^i$, and Eq.~\ref{eq:val_ind} is consistent with
Eq.~\ref{eq:harm_gf}. Though we no longer have exact conservation for fixed spacetimes,
the troublesome gauge terms in the \HARM-like formulation have been eliminated.

\section{Numerical Methods}
\label{sec:nummethods}
In this section, we provide an overview of the numerical methods we use for the
Valencia GRMHD solver. For an explanation of the refinement structure and methods shared
with other modules in \AthenaK, we refer the reader to \citet{Stone.6.24}.

\subsection{Spacetime Evolution}
\label{subsec:spacetime}
We evolve the spacetime using the Z4c formulation, which add an additional dynamical field
to the Einstein equations to damp constraint violations and allow them to propagate away
\citep{PhysRevD.81.084003,PhysRevD.88.084057}. Details of our implementation are found in
\citet{Zhu.6.24}, but we will highlight an important difference between \GRAthena and
\AthenaK which is particularly relevant to GRMHD evolution.

As noted in \citet{Daszuta:2021ecf,Cook:2023bag}, the original version of \GRAthena
evolves the spacetime on vertex-centered (VC) grids. A VC grid is convenient for
spacetime evolution on adaptive grids because the restriction operation is exact and
prolongation is only required for half the points. However, when coupled with a
cell-centered (CC) fluid scheme (as is typical for finite-volume schemes), the ADM
variables must be interpolated to cell centers for GRMHD calculations, and the matter
source terms must be interpolated to vertices for the Z4c equations. Alternatively, one
may use a CC spacetime so that the Z4c and GRMHD variables exist at the same points, but
both restriction and prolongation then require high-order interpolation operations at all
boundary points. \citet{Daszuta:2024ucu} introduced support for CC spacetimes to \GRAthena
and performed a detailed comparison between VC and CC schemes. Though both representations
have comparable accuracy, CC spacetimes offer noticeably improved performance for GRMHD
problems thanks to the lack of intergrid interpolation. Because of this computational
benefit and their simpler implementation, we have chosen to support only CC spacetimes in
\AthenaK.

\subsection{GRMHD Evolution}
\label{subsec:riemann}
We use the method of lines and standard high-resolution shock-capturing (HRSC) methods
with the Valencia solver in \AthenaK. We implement both a local Lax-Friedrichs (LLF)
scheme and the Harten-Lax-van Leer-Einfeldt (HLLE) approximate Riemann solver
\citep{Harten.1.83,EINFELDT1991273}. Though these methods are diffusive, they are
computationally efficient, guarantee positivity when coupled with an appropriate
reconstruction method, and do not typically produce unphysical results. For reconstruction
algorithms, AthenaK includes options for donor-cell reconstruction (i.e., no
reconstruction or piecewise-constant reconstruction); a piecewise-linear method (PLM) with
a van Leer monotonized slope limiter \citep{VANLEER1974361}; two variants of the
piecewise-parabolic method, including the classic method \citep{1984JCoPh..54..174C} and
a newer variant with an extrema-preserving limiter \citep{2008JCoPh.227.7069C}, which we
refer to as PPM4 and PPMX, respectively; and a fifth-order weighted essentially
non-oscillatory scheme using the Z smoothness indicator (WENOZ)
\citep{2008JCoPh.227.3191B}. We treat $\rho$, $Wv^i$, and $P$ as our primitive variables
for reconstruction. To preserve the divergence-free constraint,
$\partial_i \tilde{B}^i = 0$, \AthenaK uses the same upwind constrained transport scheme
as \AthenaPP \citep{Gardiner:2005hy,Gardiner:2007nc,Stone.7.20}. \AthenaK has several
time integrators available, and we typically use strong-stability preserving variants of
second and third-order Runge-Kutta schemes (RK2 and RK3, respectively) for GRMHD
problems \citep{Gottlieb2009}.

\subsection{Atmosphere Treatment}
\label{subsec:atmosphere}

Because a number of operations in an Eulerian code lead to division by $\rho$ or $D$,
there must be a special treatment for vacuum regions. One common solution is to add an
artificial low-density atmosphere in these regions. Though unphysical, one generally
assumes that a sufficiently rarefied atmosphere will not significantly affect the
evolution.

In practice, however, the atmosphere treatment can influence both the evolution (e.g.,
\citet{Poudel:2020fte}) and its overall stability. \AthenaK adopts a relatively simple
flooring scheme. We fill our atmosphere with a static fluid with
density $\rho_\atm$ and temperature $T_\atm$. If $\rho < f_\thr\rho_\atm$ after
reconstruction or the conserved-to-primitive inversion for some user-specified $f_\thr$,
$\rho$ and $T$ are reset to $\rho_\atm$ and $T_\atm$, respectively, the velocity is
zeroed, and $P$ is recalculated. If $T < T_\atm$, $T$ is reset and $P$ recalculated, but
both $\rho$ and $Wv^i$ are left alone. A similar policy holds for the conserved variables.
For given values of $D$ and $B^i$, a floor $\tau_\atm(D,B^i)$ assuming $T=T_\atm$ can be
calculated. If $\tau < \tau_\atm$, $\tau$ is reset to $\tau_\atm(D,B^i)$. If
$D < f_\thr \rho_\atm$, we set $D = \rho_\atm$, $S_i = 0$, and
$\tau = \tau_\atm(D_\atm,B^i)$. In all cases, $B^i$ is left alone to preserve the
divergence-free condition. However, if $B^2/D > M_\tmax$, we reset $D = B^2/M_\tmax$.

Though this procedure generally allows for stable evolutions, it artificially injects mass
and heat into the fluid and removes momentum from the system. To help address this, both
the Valencia and \HARM-like GRMHD solvers use a first-order flux correction (FOFC)
\citep{Lemaster:2008gh}. After the flux calculation, the next Runge-Kutta step is
estimated. If any cell in the new state requires a floor or the conserved-to-primitive
inversion (see \ref{subsec:c2p}) fails, it is flagged. We then recalculate the fluxes for
all flagged cells using a more dissipative scheme; namely, we reduce reconstruction to
first order (donor-cell reconstruction) and recalculate the fluxes using LLF or HLLE. This
scheme is guaranteed to be positivity-preserving for $D$ (and $\tau$ in flat space). While
this does not eliminate the need for a floor or even ensure that $D$ cannot fall below the
floor (e.g., rarefaction waves in strong explosions), it does prevent floors caused by
spurious oscillations near discontinuities, such as stellar surfaces or shocks, and it
also adds a small amount of dissipation where a state might become unphysical. We also
stress that FOFC is only an estimate; we do not apply source terms, perform constrained
transport, or update the matter fields prior to testing the solution.

One shortcoming of the proposed FOFC method is that it only identifies unphysical states
by failures during the primitive variable recovery. However, unphysical behavior, such
as carbuncles or spurious oscillations, may develop over several time steps before the
conserved-to-primitive inversion fails. Attempting to fix these errors after the fact
using FOFC may prevent the inversion failure, but it does not correct the unphysical
state itself.

In an attempt to address this problem, consider the following: scalar conservation
laws obey a maximum principle \citep{Kruzkov_1970}. That is, for
\begin{equation}
  \label{eq:scalcons}
  \partial_t u + \partial_i f^i(u) = 0,
\end{equation}
it follows that for $u(x,t)$,
\begin{equation}
  \label{eq:scaldmp}
  \min_x u(x,0) \leq u(x,t) \leq \max_x u(x,0).
\end{equation}
When this equation is discretized, such as in a finite-volume scheme, a similar notion
holds, known as the local discrete maximum principle (DMP). The DMP does not strictly hold
for a system of conservation laws like the GRMHD equations, but we can use it as a
starting point for a scheme which limits excessive growth. Let $U^n_i$ represent a state
at time $t = n\Delta t$ averaged over the cell centered on position $x_i$. We also define
$\mathcal{V}_i$ as the set containing the cell at $x_i$ and all its immediate neighbors,
including corner cells. For $M \geq 1$, we require that
\begin{equation}
  \label{eq:vecdmp}
  \frac{1}{M}\min_{j\in \mathcal{V}_i} U^n_j \leq U^{n+1}_i \leq 
      M\max_{j\in \mathcal{V}_i} U^n_j.
\end{equation}
We may consider this a sort of \textit{relaxed} DMP. The GRMHD equations are not required
to obey the relaxed DMP, either, but for an appropriately chosen $M\geq 1$, applying the
FOFC where it is violated will add additional dissipation to help limit spurious
oscillations or numerical instabilities. In the tests which follow, we choose $M$ somewhat
arbitrarily (typically $M=1.1$ or $M=1.2$) to damp large oscillations while still allowing
some growth. Detailed analysis or numerical experiments may reveal an optimal value for
a given problem or set of equations, but this is beyond the scope of this work. Our own
implementation also only enforces the DMP on $D$ and $\tau$, as enforcing the maximum
principle on $S_i$ is both unphysical and adds excessive dissipation. Furthermore, we
perform this step \textit{before} the $\tau$ source terms are added during each substep of
the time integration in order to avoid mistaking gravitational effects for physically
spurious solutions. We refer to this method throughout the paper as FOFC+DMP. Note that
this approach is not novel; the FOFC+DMP method we propose here may be considered a
simplified version of a multi-dimensional optimal-order detection (MOOD) scheme, which
selects the highest-order method satisfying physical constraints and an appropriate
maximum principle \citep{CLAIN20114028,Zanotti:2015mia}.

\begin{table*}[!htb]
\begin{ruledtabular}
  \caption{\label{tab:shocktubes} Initial conditions for the Balsara shock tube tests,
  consisting of a Riemann problem with discontinuity located at $x=0.5$ with left ($L$)
  and right ($R$) states.}
  \begin{tabular}{ll|ccccccccc}
    Test & State & $\Gamma$ & $\rho$ & $P$ & $v^x$ & $v^y$ & $v^z$ & $B^x$ & $B^y$ & $B^z$ \\
    \hline
    1 & $L$ & $2$ & $1$ & $1$ & $0$ & $0$ & $0$ & $0.5$ & $1$ & $0$ \\
      & $R$ &     & $0.125$ & $0.1$ & $0$ & $0$ & $0$ & $0.5$ & $-1$ & $0$ \\
    \hline
    2 & $L$ & $5/3$ & $1$ & $30$ & $0$ & $0$ & $0$ & $5$ & $6$   & $6$ \\
      & $R$ &       & $1$ & $1$  & $0$ & $0$ & $0$ & $5$ & $0.7$ & $0.7$ \\
    \hline
    3 & $L$ & $5/3$ & $1$ & $10^3$ & $0$ & $0$ & $0$ & $10$ & $7$ & $7$ \\
      & $R$ &       & $1$ & $0.1$  & $0$ & $0$ & $0$ & $10$ & $0.7$ & $0.7$ \\
    \hline
    4 & $L$ & $5/3$ & $1$ & $0.1$ & $0.999$  & $0$ & $0$ & $10$ & $7$  & $7$ \\
      & $R$ &       & $1$ & $0.1$ & $-0.999$ & $0$ & $0$ & $10$ & $-7$ & $-7$ \\
    \hline
    5 & $L$ & $5/3$ & $1.08$ & $0.95$ & $0.4$   & $0.3$  & $0.2$ & $2$ & $0.3$  & $0.3$ \\
      & $R$ &       & $1$    & $1$    & $-0.45$ & $-0.2$ & $0.2$ & $2$ & $-0.7$ & $0.5$
  \end{tabular}
\end{ruledtabular}
\end{table*}

\subsection{Primitive Variable Recovery}
\label{subsec:c2p}
A well-known difficulty with GRMHD concerns the conserved-to-primitive inversion problem.
Though Eq.~\ref{eq:conserved} provides a simple closed-form expression for the conserved
variables in terms of the primitive variables, the inverse operation is transcendental
for most equations of state (EOSs). Our approach to the problem is based on the algorithm
presented in \citet{2021PhRvD.103b3018K}. Though a reference library, \RePrimAnd, exists
for this problem \citep{RePrimAnd.2021}, we have implemented our own library,
\PrimitiveSolver \footnote{see \url{https://github.com/jfields7/primitive-solver}}, for
use in \GRAthena and \AthenaK \citep{Cook:2023bag}. The primary reason for this decision
is that \RePrimAnd extensively uses polymorphism and virtual functions, which are more
challenging to use inside GPU kernels.

\PrimitiveSolver uses a templated policy-based design which mimics many of the traditional
features of polymorphism. A base class \code{EOS}\xspace depends on two template
parameters, \code{EOSPolicy}\xspace and \code{ErrorPolicy}\xspace, which provide the
particular EOS implementation and error response (including the atmosphere treatment).
Unlike traditional polymorphism, however, \code{EOS}\xspace inherits directly from both
\code{EOSPolicy}\xspace and \code{ErrorPolicy}\xspace, rather than \code{EOS}\xspace
holding pointers to abstract policy classes whose concrete derived classes implement
these features. Therefore, customizable EOS or error calls can be made on the GPU because
the appropriate template functions are resolved at compile time rather than via a virtual
method table at runtime.

The version of \PrimitiveSolver included in \AthenaK supports ideal gases and piecewise
polytropic hybrid EOSs, and microphysical tabulated EOSs are currently being tested. For
more information, including algorithmic differences from \RePrimAnd, we refer the reader
to the appendix of \citet{Cook:2023bag}.

\section{Tests}
\label{sec:tests}
In this section, we describe a number of tests to validate \AthenaK.

\subsection{Magnetized Shock Tubes}
\label{subsec:shocktubes}
As a first test, we consider a set of magnetized shock tubes in one dimension, following
the standard setups in flat space presented by \citet{Balsara_2001}. We summarize the
initial conditions in Table~\ref{tab:shocktubes}. Each shock tube consists of a  Riemann
problem with constant left and right states separated by a discontinuity at $x=0.5$ on a
domain of $[0,1]$.

\begin{figure*}
  \includegraphics{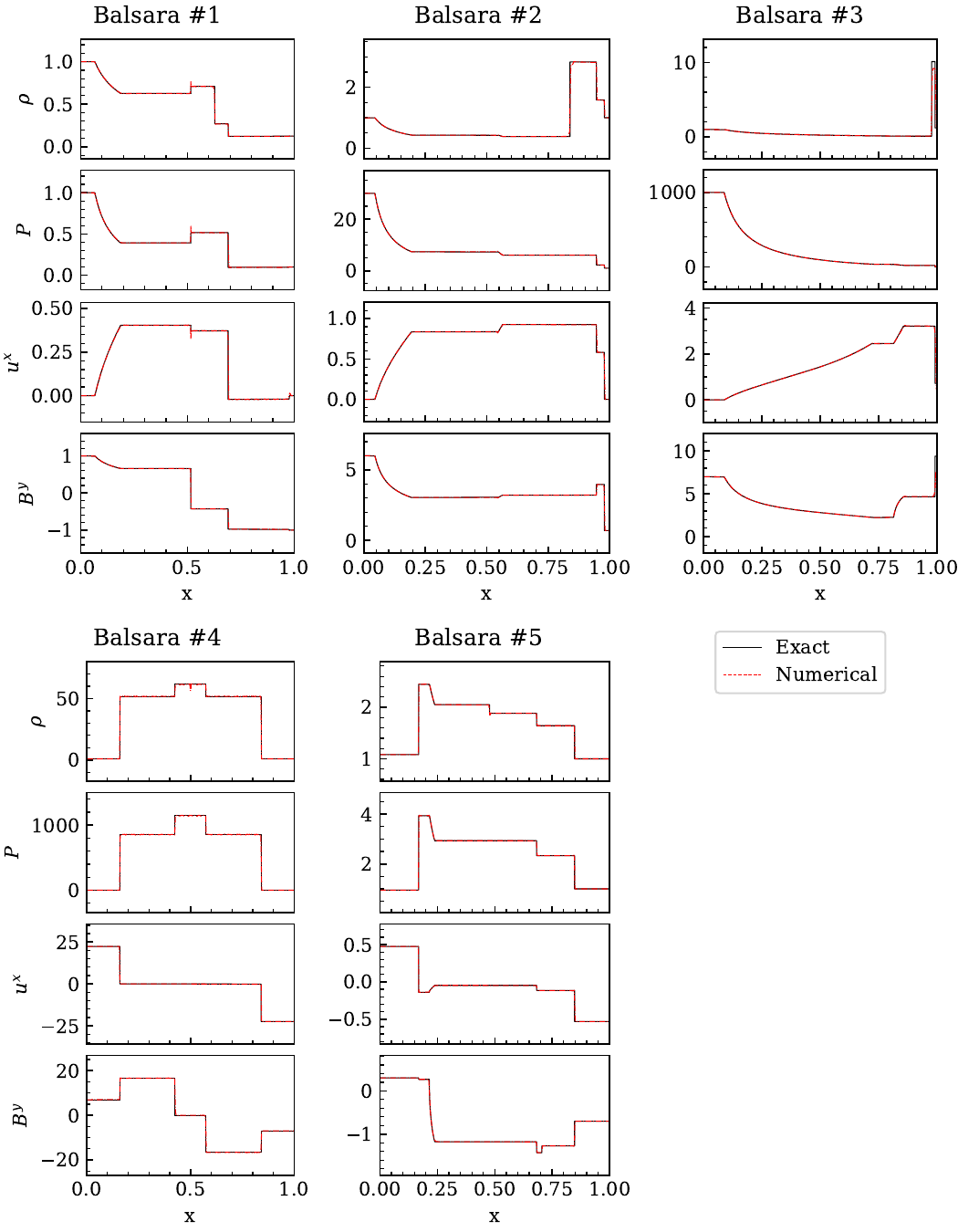}
  \caption{\label{fig:shock_tubes} The rest-mass density $\rho$, pressure $P$,
    $x$-velocity $u^x=Wv^x$, and $y$-component of the magnetic field $B^y$ for the shock
    tube tests at $t=0.5$.}
\end{figure*}

%
%
%
%

Our computational domain contains $1600$ cells. We use the LLF Riemann solver with PPM4
reconstruction. Because the fluid solver remains stable during this test and never
approaches atmosphere, we do not enable the FOFC scheme. We evolve these systems to
$t=0.5$ and compare them to their corresponding exact solutions \citep{Giacomazzo:2005jy}
in Figure~\ref{fig:shock_tubes}. For brevity, we only show the
$x$-component of velocity and the $y$-component of the magnetic field. The results are in
good agreement with the exact solution.

\begin{figure}
  \includegraphics[width=\columnwidth]{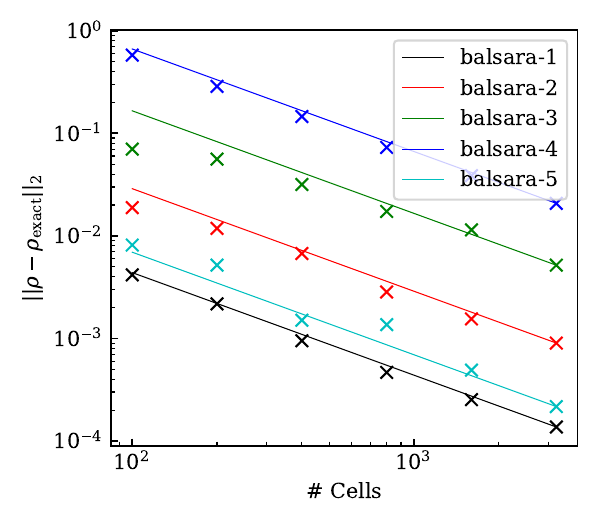}
  \caption{\label{fig:shock_conv} Shock tube convergence as measured by the $L^2$ error
  between the numerical and exact solutions for $\rho$ at $t=0.5$. Markers indicate
  numerical solutions, and solid lines measure first-order convergence.}
\end{figure}

We also perform additional tests with 100, 200, 400, 800, and 3200 cells in order to
estimate convergence as measured by the $L^2$ norm of the error in density. We show these
results in Figure~\ref{fig:shock_conv}. Some of the lower-resolution solutions do not
appear to be in a completely convergent regime, most notably for Balsara Tests \#3 and
\#5. However, we do achieve approximate first-order convergence for all tests, as
expected for solutions dominated by shocks.

\subsection{Magnetized Cylindrical Blast Wave}
\label{subsec:blast}
Our second test considers a magnetized cylindrical blast wave in flat spacetime following
\citet{1999MNRAS.303..343K}. The domain is filled with a static low-density,
low-pressure fluid of $\rho_\textrm{amb}$ and $P_\textrm{amb}$. Inside a specified inner
radius, $r_\textrm{in}$, we set $\rho_\textrm{in}$ and $P_\textrm{amb}$. From
$r_\textrm{in}$ out to an outer radius $r_\textrm{out}$, $\rho$ and $P$ decay
exponentially. The magnetic field is set to $B^i = \left(B_x,0,0\right)$.

For this test, we set $\rho_\textrm{amb} = 10^{-4}$, $P_\textrm{amb}=3\times10^{-5}$,
$\rho_\textrm{in} = 10^{-2}$, $P_\textrm{in} = 1$, $r_\textrm{in} = 0.8$, and
$r_\textrm{out} = 1.0$. We consider three values for the magnetic field: $B_x = 0.01$,
$B_x = 0.1$, and $B_x = 1.0$. The EOS is an ideal gas with $\Gamma = 4/3$. As noted in
\citet{Beckwith:2011iy}, this test is particularly difficult for large $B_x$ because the
combination of relatively high Lorentz factors and high magnetizations heavily often
causes the conserved-to-primitive inversion to return poor-quality solutions (if it finds
them at all). Common solutions to this issue involve altering the test (e.g., increasing
the ambient pressure as in \citet{Leisman2005,DelZanna:2007pk,Beckwith:2011iy,
Deppe:2023qxa}) or employing schemes which violate energy conservation
\citep{1999MNRAS.303..343K,Mignone:2006jg,Phillips:2023shg,Komissarov:2024kva}.
Consequently, the original setup as in \citet{1999MNRAS.303..343K} tests both the
robustness of our primitive recovery routine and the ability of the FOFC and FOFC+DMP
schemes to add enough dissipation to limit unphysical behavior.

The computational domain is a 2D Cartesian grid spanning $[-6.0, 6.0]$ with $200$ cells
in each direction. We employ HLLE as our Riemann solver, use WENOZ for reconstruction, and
evolve in time with RK3 using $\textrm{CFL}=0.1$. We set the floor to $\rho_\atm=10^{-10}$
and $T_\atm=10^{-8}$ with $f_\thr = 1.0$. For the FOFC+DMP run, we set $M=1.1$.

\begin{figure}
  \includegraphics[width=\columnwidth]{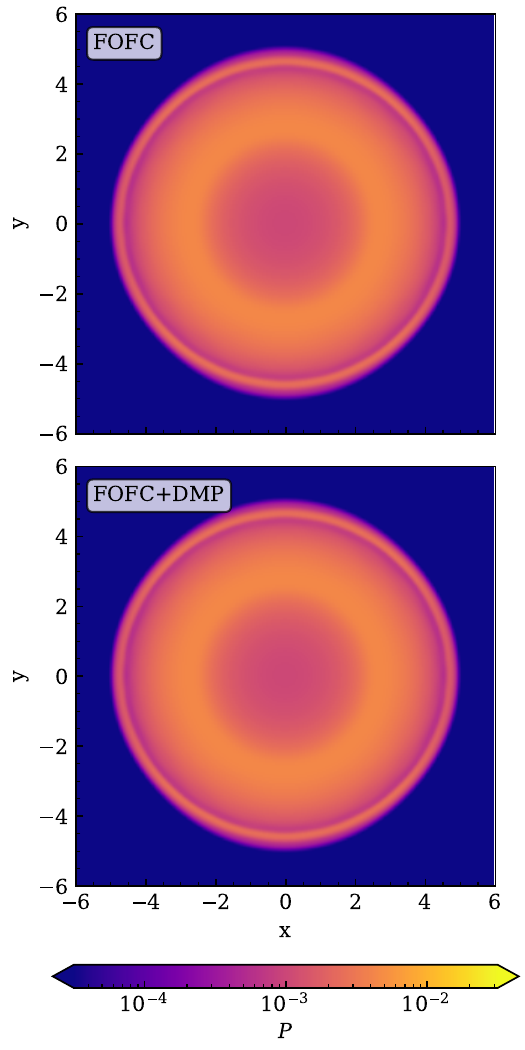}
  \caption{\label{fig:bx1e-2}The pressure of a weakly magnetized blast wave ($B_x = 0.01$)
      at $t=4$ for the FOFC (top) and FOFC+DMP (bottom) methods.}
\end{figure}

\begin{figure}
  \includegraphics[width=\columnwidth]{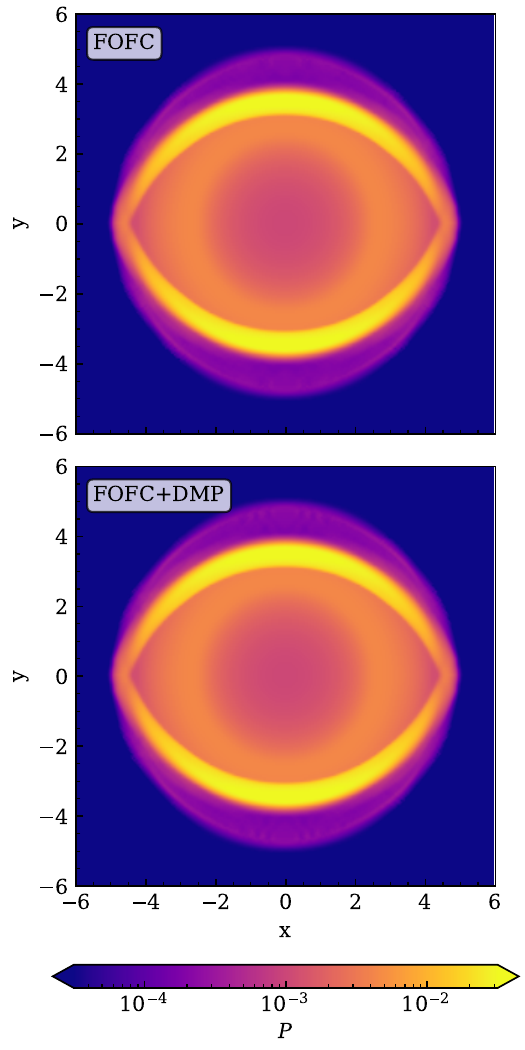}
  \caption{\label{fig:bx1e-1}The same as Figure~\ref{fig:bx1e-2}, but for a moderately
      magnetized blast wave ($B_x=0.1$)}
\end{figure}

\begin{figure}
  \includegraphics[width=\columnwidth]{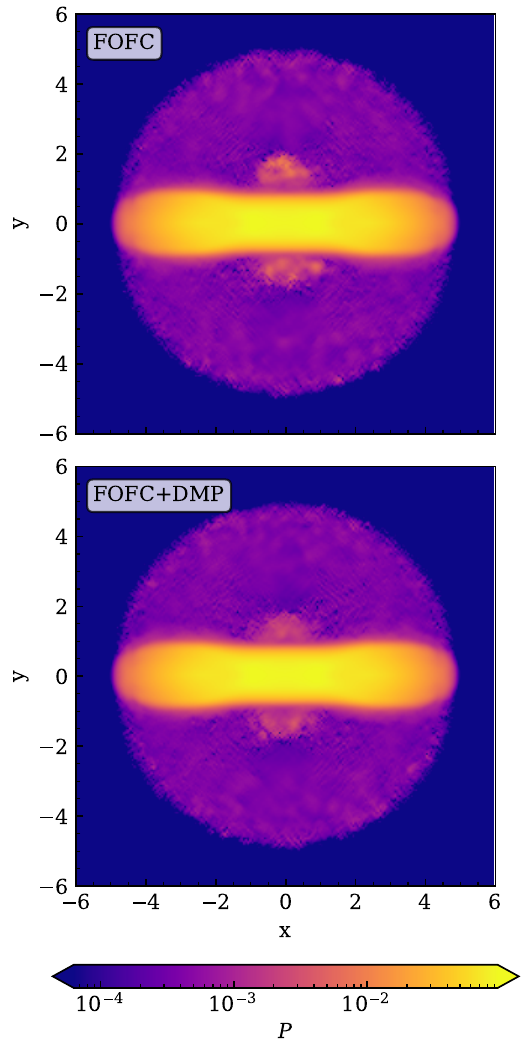}
  \caption{\label{fig:bx1}The same as Figure~\ref{fig:bx1e-2}, but for a strongly
      magnetized blast wave ($B_x=1.0$)}
\end{figure}

We show the results for the $B_x = 0.01$ case in Figure~\ref{fig:bx1e-2}, the $B_x = 0.1$
case in Figure~\ref{fig:bx1e-1}, and $B_x = 1.0$ in Figure~\ref{fig:bx1}. Both the FOFC
and FOFC+DMP schemes are in good agreement for the weakly and moderately magnetized cases.
The strongly magnetized case exhibits significant noise for both solutions, with FOFC+DMP
offering somewhat better behavior near the center. Nevertheless, the overall solution
agrees with \citet{1999MNRAS.303..343K}, and our results are consistent with what was
found for the same test using the HARM-like solver \citep{Stone.6.24}.

We also ran tests of the strongly magnetized case without FOFC (not pictured) to validate
our claims of added stability; a test with PLM did not resolve the bar feature in
Figure~\ref{fig:bx1} as clearly, and the surrounding blast wave showed strong
instabilities which rendered the solution useless. A test with WENOZ quickly disintegrated
due to primitive inversion failures resetting the solution to atmosphere. With adjustments
to the atmosphere and the error policy, it is possible that we could have improved
stability in the non-FOFC solution. However, these are only stopgap measures which do not
address the real problem: the solution is becoming unphysical. The FOFC and FOFC+DMP
methods represent simple ways to add dissipation where the worst instabilities would
develop while using a more accurate scheme in better-behaved regions.

\subsection{Magnetic Loop Advection}
\label{subsec:loop_advection}
This next test considers a two-dimensional magnetic loop which is advected at constant
velocity \citep{DEVORE1991142}. As in \citet{Gardiner:2005hy,Beckwith:2011iy}, the
magnetic field components $B^{x}$ and $B^{y}$ are calculated from a vector potential,
\begin{equation}
\label{eq:loop_A}
  A_z = \begin{cases}
          A_0 (R - r) & r \leq R \\
          0           & r > R,
        \end{cases}
\end{equation}
where $r = \sqrt{x^2 + y^2}$. Our setup most closely resembles \citet{Deppe:2023qxa},
where we fix $A_0 = 10^{-3}$, $R = 0.3$, $\rho = 1$, $P = 3$, and
$\mathbf{v} = \{1/1.2,1/2.4,0\}$ on a periodic domain covering
$[-0.5,0.5]\times[-0.5,0.5]$ with $240^2$ cells. For this test, we use RK3, HLLE+FOFC, and
WENOZ and evolve a single period ($t=2.4$) with a CFL of $0.25$.

\begin{figure}
  \includegraphics[width=\columnwidth]{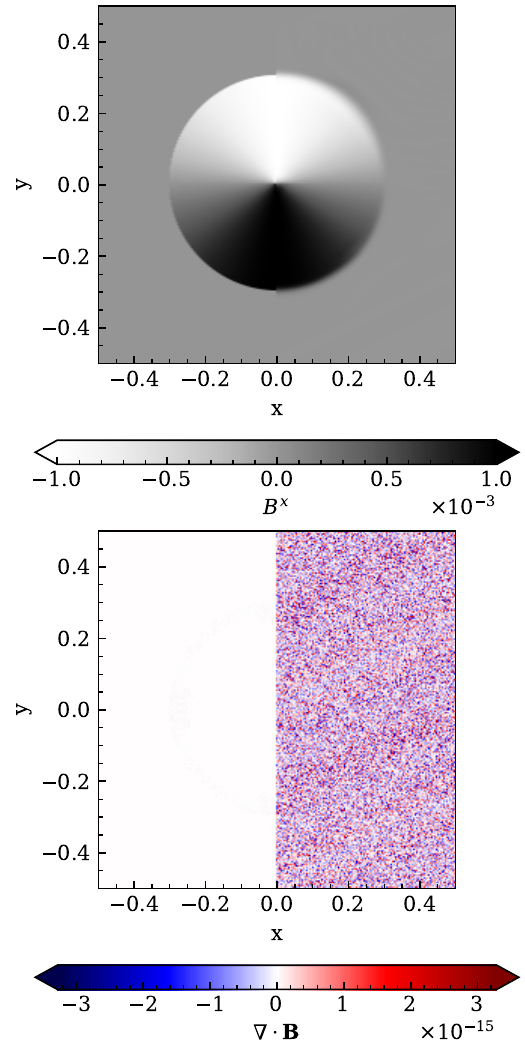}
  \caption{\label{fig:loop}(Top) The $B^x$ component of the magnetic field during the loop
    advection test. (Bottom) The violation of the divergence-free constraint relative to
    the maximum magnetic field strength. In both plots, the left half shows the initial
    data, and the right half shows the evolved data after a single period ($t=2.4$).}
\end{figure}

This test probes the correctness and the diffusiveness of the constrained-transport
scheme. As shown in Figure~\ref{fig:loop}, \AthenaK advects the loop at the correct speed.
Some numerical diffusion is apparent, but the shape is correct with no obvious artifacts.
Maximum relative errors in the divergence-free condition are around $7\times10^{-14}$ and
primarily concentrated around the edge of the loop. After a single period, the maximum
error has amplified to approximately $3\times10^{-12}$, which is consistent with growth
from floating-point errors.

\subsection{Oscillating Neutron Star in the Cowling Approximation}
\label{subsec:cowling_ns}
Our first test in curved spacetime considers linear perturbations to a
Tolman-Oppenheimer-Volkoff (TOV) star. Our test setup is designed to replicate the
isolated neutron star tests in \citet{Radice:2013xpa} as closely as possible. For these
tests, we adopt HLLE as our Riemann solver with PPM4 reconstruction.

The computational domain spans $[0~\Msun, 102.4~\Msun]$ in all dimensions with vacuum
boundary conditions and a reflection symmetry along the $x$, $y$, and $z$ axes. The base
refinement level has $128$ cells in each direction, and we use static mesh refinement
(SMR) to refine all mesh blocks inside $51.2~\Msun$ and then again at $25.6~\Msun$ to give
the innermost refinement region a resolution of $0.2~\Msun$. We set the
Courant-Friedrich-Lewy (CFL) factor to $\textrm{CFL}=0.4$.

The TOV solver uses the polytrope $P=K \rho^\Gamma$ with $K = 100$, $\Gamma = 2$, and
central density $\rho_c = 1.28\times10^{-3}~\Msun^{-2}$, which results in a single star
with mass $M = 1.4~\Msun$ and radius (in Schwarzschild coordinates) of
$R\approx 9.59~\Msun$. We also perform our test in the Cowling approximation, i.e., with
spacetime evolution disabled, and we do not add a magnetic field. To match the setup in
\citet{Radice:2013xpa}, we set $\rho_\atm = 10^{-10}~\Msun^{-2}$ and $f_\thr = 1.01$.
Because we evolve the system with an ideal gas, we also add a temperature floor of
$T=10^{-8}~\Msun$.

To induce a small perturbation in the velocity, we use the radial velocity eigenfunction
of \citet{Novak:2001ck,Noble:2007vf},
\begin{equation}
  \label{eq:vel_prof}
  v_r(r) = \frac{U}{2}(3x - x^3),
\end{equation}
with $x \equiv r/R$ and $U$ indicating the amplitude of the profile at stellar surface.
For our tests, we use $U = -0.024$, representing an inward perturbation.

\begin{figure}
  \includegraphics[width=\columnwidth]{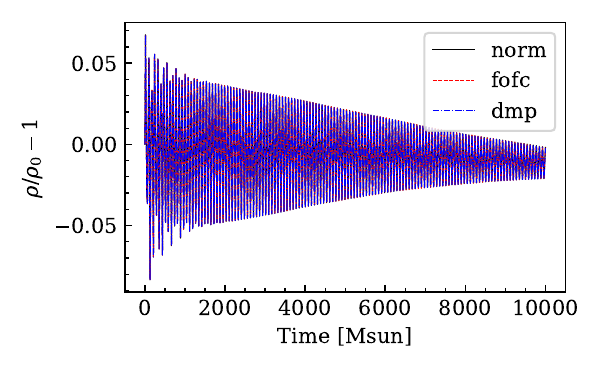}
  \includegraphics[width=\columnwidth]{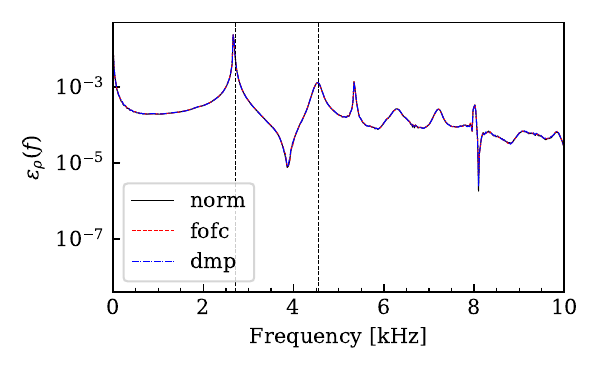}
  \caption{\label{fig:tov-time} Oscillations of the central density of the TOV star as a
    function of time (top) and the power spectrum of central density oscillations
    (bottom). The 'norm', 'fofc' and 'dmp' labels refer to runs without FOFC, with FOFC,
    and with FOFC+DMP, respectively. To reduce spectral leakage, the oscillations were
    filtered with a Tukey window of $\alpha=0.203$ prior to computing the Fourier
    transform. The black vertical lines mark the fundamental frequency and its first
    harmonic as predicted by perturbation theory.}
\end{figure}

We plot the central density oscillations in time and frequency space in
Figure~\ref{fig:tov-time}. All three methods also show excellent agreement in oscillations
of the central density; the power spectrum of these oscillations also show that all three
methods differ from the peak frequency and its first harmonic as predicted by perturbation
theory by $47~\Hz$ and $20~\Hz$, respectively \citep{Stergioulas:2003ep}. The frequency
resolution is ${\sim}20~\Hz$, suggesting that the error is small even at this relatively
coarse resolution. Nevertheless, the low resolution does cause a slow downward drift in
the central density. The similarity of the results indicates that
the FOFC and FOFC+DMP methods have minimal effects on stability, accuracy, and damping
time. This is expected, as these methods should only trigger to prevent primitive
inversion failures or unnecessary flooring.

\subsection{Free Evolution of an Oscillating Neutron Star}
\label{subsec:free_ns}
To test the spacetime evolution, we consider the same test as \ref{subsec:cowling_ns} but
with all symmetry removed and the Z4c solver enabled. We also do not apply an explicit
perturbation and instead rely on discretization errors to excite oscillations. To improve
gauge stability, we switch to isotropic coordinates, which slightly reduces the coordinate
radius to $R\approx8.13~\Msun$. We extend the boundary out to $\pm204.8~\Msun$ in all
directions and add four refinement levels, each successively halving the domain, such that
the finest level spans $\pm12.8~\Msun$ in each direction. We perform tests with $192^3$
and $384^3$ cells on the base grid, which corresponds to resolutions of
$\Delta x \approx 0.133~\Msun$ ($\Delta x \approx 197~\mathrm{m}$) and
$\Delta x \approx 0.0667~\Msun$ ($\Delta x \approx 98~\mathrm{m}$) on the finest
refinement level. We additionally change the atmosphere to
$\rho_\atm=1.28\times10^{-13}$, $T_\atm=1.28\times10^{-11}$, and $f_\thr = 0.1$ (i.e., the
density is allowed to drop below the atmosphere). To improve accuracy near the center of
the star, we switch the reconstruction algorithm to WENOZ. Like the Cowling case, we
perform tests with no FOFC, with FOFC, and with FOFC+DMP.

\begin{figure}
  \includegraphics[width=\columnwidth]{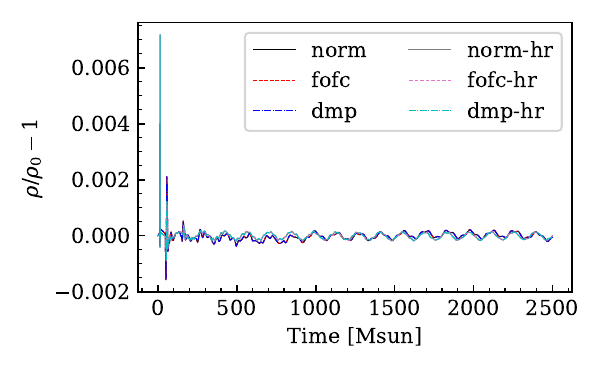}
  \includegraphics[width=\columnwidth]{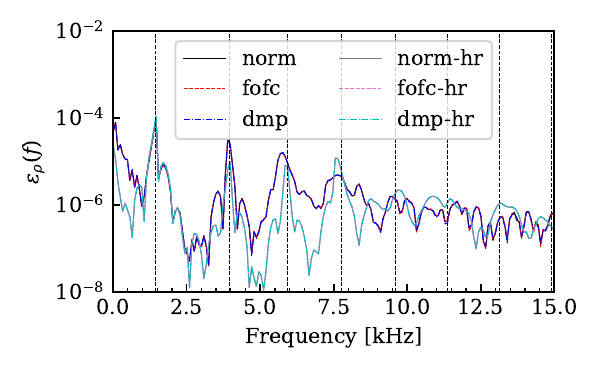}
  \caption{\label{fig:free-tov} The same as Figure~\ref{fig:tov-time}, but for a TOV star
  with the Z4c solver enabled. The `norm', `fofc', and `dmp' labels refer to runs without
  FOFC, with FOFC, and with FOFC+DMP, respectively. Labels without the `-hr' suffix refer
  to the $192^3$ runs, and those with to the $384^3$ solutions.}
\end{figure}

Figure~\ref{fig:free-tov} shows the central density oscillations and their power spectrum.
Following an initial spike as the TOV solution settles onto the computational grid, the
oscillations remain quite small without strong damping or indications of secular drift.
As in the Cowling approximation, all three methods demonstrate good agreement with the
oscillation frequencies predicted by perturbation theory \citep{Yoshida:1999vj}. The
$192^3$ solution has errors of $19~\Hz$, $56~\Hz$, and $149~\Hz$ compared to the predicted
values for the fundamental and the first and second harmonics, respectively. The errors of
both the fundamental and the first harmonic are less than the frequency resolution
(${\sim}81~\Hz$). The third harmonic also seems to be present and has a fairly low error
around $141~\Hz$, but the width and relatively low amplitude of this peak suggest it may
not be reliable. The $384^3$ solution has errors of $19~\Hz$, $25~\Hz$, $68~\Hz$, and
$222~\Hz$ for the fundamental frequency and the first, second, and third harmonics. The
third harmonic is the first with an error exceeding the finite frequency resolution. The
fourth harmonic has an error of $157~\Hz$, though this peak, as well as those which
follow, may not be reliable.

\subsection{Magnetized Accretion Disk}
\label{subsec:fm_disk}
In this test, we simulate the development of the magnetorotational instability (MRI)
inside an accretion disk in a fixed spacetime. We use initial conditions which are
similar to the Event Horizon Telescope (EHT) code comparison test
\citep{EventHorizonTelescope:2019pcy}, which consists of a Fishbone-Moncrief torus
\citep{1976ApJ...207..962F} surrounding a Kerr black hole of mass $M$ with dimensionless
spin $a=0.9375$. The torus begins at an inner radius $r_\mathrm{in}=6M$ and extends to an
outer radius $r_\mathrm{out}=12M$ with a magnetic field defined by the vector potential
\begin{equation}
  \label{eq:eht_A}
  A_\phi = A_0 \max\left(\rho/\rho_\mathrm{max}-0.2,0\right),
\end{equation}
where $A_0$ is set so that $\beta=2P/b^2=100$. We assume an ideal gas with $\Gamma=4/3$,
and we add random perturbations to the pressure which are uniformly distributed in the
range $[-0.02P,0.02P]$ to help excite the MRI. Outside the torus, the background density
and pressure are set to $\rho=1.0\times10^{-5}r^{-3/2}$ and $P=3.33\times10^{-8}r^{-5/2}$
in code units, and we choose flooring parameters $\rho_\mathrm{atm}=10^{-10}$,
$P_\mathrm{atm}=3.33\times10^{-13}$ (for the \HARM-like formulation), and
$T_\mathrm{atm}=3.33\times10^{-13}$ (for the Valencia formulation) in code units.

We evolve the torus with both the new Valencia GRMHD solver and the existing \HARM-like
solver with RK2, HLLE+FOFC, and PPM4 to $t=10000M$ on a domain spanning $\pm64M$ in each
direction with four levels of refinement, each successively halving the size of the grid.
Each system is run at three resolutions; the base grid has $64^3$,
$96^3$, and $128^3$ cells, and the mesh blocks have a width of $16$, $24$, and $32$ cells,
respectively. This leads to resolutions of $\Delta x_\mathrm{64} = 0.125M$,
$\Delta x_\mathrm{96} = 0.0833M$, and $\Delta x_\mathrm{128} = 0.0625M$ on the finest
refinement level. We excise the region $r\leq M$ and fill it with atmosphere to improve
stability inside the horizon. The magnetic field is not modified during excision.

The purpose of this test is two-fold: first, it validates the Valencia GRMHD solver in
curved space by showing it is in agreement with the \HARM-like solver. Because the two
solvers are physically equivalent, we expect that they should converge to the same
results.

Secondly, it allows us to compare the strengths and weaknesses of these different
formulations. As shown in Sec.~\ref{sec:mhd}, the \HARM-like and Valencia formulations
treat the energy evolution differently, so we expect that they should display different
numerical properties. In particular, by evolving $T^0_0$ instead of $E$, the \HARM-like
formulation should exactly conserve the fluid energy in any stationary spacetime. This
may also lead to a more stable evolution in the strong-field regime, where $E$ or $\tau$
in the Valencia formulation can develop large source terms which may cause issues with
round-off error or preserving positivity. However, it is not immediately clear how
important these effects are in a full nonlinear evolution.

Though some of the codes used in the EHT code comparison test, including \ECHO
\citep{DelZanna:2007pk}, \BHAC \citep{Porth:2016rfi}, and \IllinoisGRMHD
\citep{Etienne:2015cea}, use the Valencia formulation and achieve similar results to codes
using a \HARM-like formulation, there are differences in grid setups and numerical methods
which make a one-to-one comparison difficult. \citet{Porth:2016rfi} provide a more
consistent comparison and find good agreement, but their analysis is limited to the mass
accretion rate and magnetic flux in a 2D axisymmetric problem.

To compare the consistency in evolution, we use the following radial fluxes as
diagnostics:
\begin{subequations}
\begin{align}
  \label{eq:Mdot}
  \dot{M} &= \intrad{\rho u^r \sqrt{-g}}, \\
  \label{eq:Edot}
  \dot{E} &= \intrad{-T^r_t\sqrt{-g}}, \\
  \label{eq:Ldot}
  \dot{L} &= \intrad{T^r_\phi\sqrt{-g}}, \\
  \label{eq:mag_phi}
  \Phi_\mathrm{B} &= \frac{1}{2}\intrad{\left|b^r u^0 - b^0 u^r\right|\sqrt{-g}},
\end{align}
\end{subequations}
where we integrate over a sphere with a fixed Kerr-Schild radius $r$. In practice this
means the integration surface is oblate in the usual Cartesian coordinates, and we
interpolate our primitive variables to an appropriate geodesic grid first.

\begin{figure*}
  \includegraphics{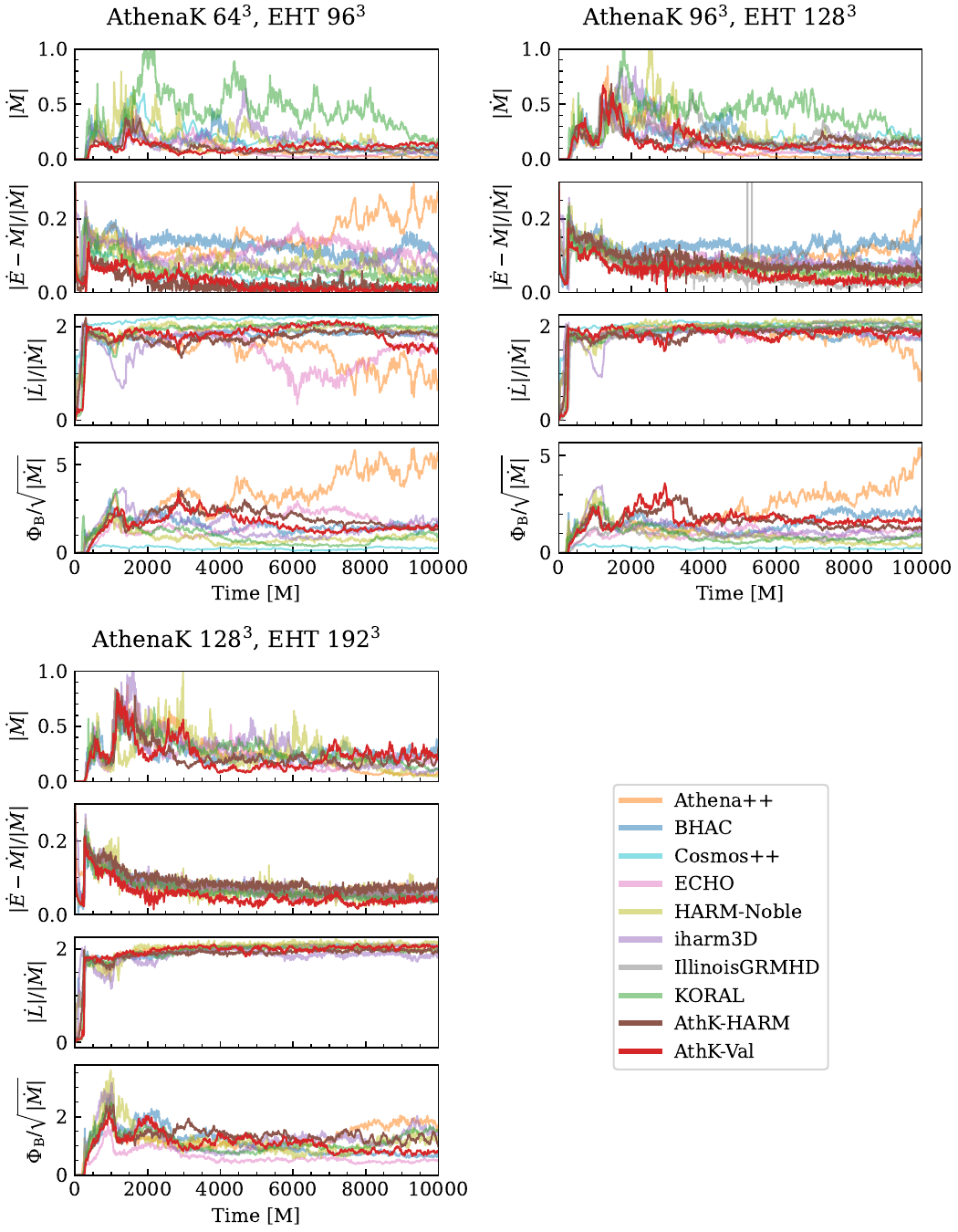}
  \caption{\label{fig:eht_all} Diagnostic fluxes for the $64^3$, $96^3$, and $128^3$ cases
    plotted against some of the $96^3$, $128^3$, and $192^3$ EHT data (respectively).
    From top to bottom in each set of plots, we show the rates for mass accretion,
    normalized energy accretion, normalized angular momentum accretion, and normalized
    magnetic flux.}
\end{figure*}

%
%

\begin{figure*}
  \centering
  \includegraphics{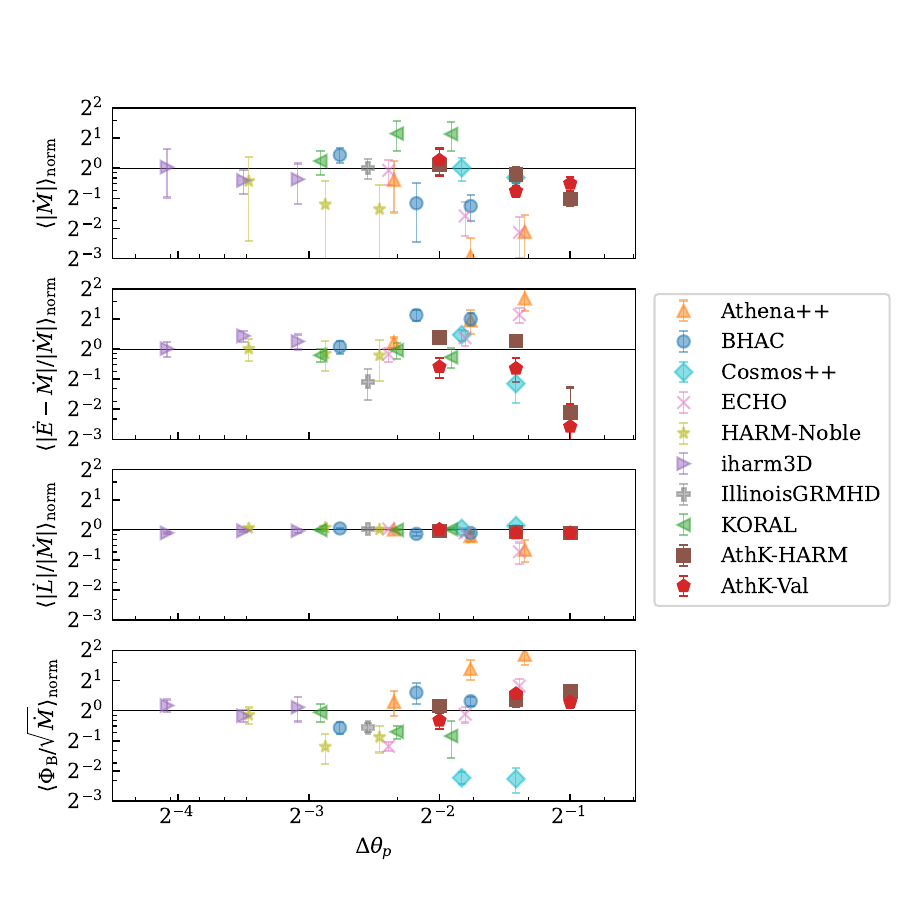}
  \caption{\label{fig:eht_avg} The average of the diagnostic fluxes over
      $t\in[5000M,10000M]$ for each resolution, normalized by the average of all the codes
      considered. The error bars span a single standard deviation.}
\end{figure*}

We show these diagnostic quantities in Figure~\ref{fig:eht_all} for each run. For
comparison, the plots also contain a subset of the data from
\citet{EventHorizonTelescope:2019pcy} (hereafter referred to as the EHT data), namely
\AthenaPP, \BHAC, \CosmosPP \citep{Anninos:2005kc,2012ApJS..201....9F,Fragile:2014bfa},
\ECHO, \HARMNoble (usually called \code{HARM3d}\xspace) and \iharm
\citep{Gammie:2003rj,Noble:2005gf,Noble:2008tm}, \IllinoisGRMHD, and \KORAL
\citep{2013MNRAS.429.3533S,Sadowski:2013gua}. Somewhat confusingly, we compare the $64^3$
\AthenaK runs to the $96^3$ EHT data, the $96^3$ runs to the $128^3$ EHT data, and the
$128^3$ runs to the $192^3$ EHT data. This is because the \AthenaK grids do not have a
direct correspondence with any setup in the EHT data; all \AthenaK setups are Cartesian,
while most of the EHT data setups are on logarithmic spherical grids. Nevertheless, we see
good agreement between \AthenaK and the other codes present at all resolutions,
particularly for the angular momentum flux and the mass accretion rates.

To examine this more carefully, Figure~\ref{fig:eht_avg} plots the averages of each of the
diagnostic fluxes over the range $t\in[5000M,10000M]$, normalized to the total average for
all the codes considered, as a function of resolution. Because most of the compared codes
use spherical coordinates and differing grid setups, we follow
\citet{EventHorizonTelescope:2019pcy} and define a fiducial resolution,
$\Delta \theta_p \equiv \sqrt{g_{\theta\theta}(12M,\pi/2)}\Delta \theta$, corresponding to
the proper distance between polar grid cells at $r=12M$ in the equatorial plane. We treat
$\Delta z$ as the equivalent quantity for the Cartesian runs. As seen in the time series
plots, \AthenaK is consistent with the expected results based on the EHT data, with the
only true outlier being the energy accretion in the $64^3$ test.

The $64^3$ \AthenaK runs do display a much lower value for $|\dot{E}-\dot{M}|/|\dot{M}|$
than other codes, but this appears to be an artifact of the resolution; both the $96^3$
and $128^3$ runs have energy accretion rates which are more consistent with other GRMHD
codes. This is most likely because of the grid setup we have chosen. Though we have
reasonable resolution near the horizon, with our $64^3$ setup exceeding the resolution of
the \IllinoisGRMHD data, the resolution quickly falls off in the torus to
$\Delta z = 0.5M$ at $(12M,0,0)$, which is lower than the fiducial resolution in any of
the other codes.

When compared to each other, the \HARM-like and Valencia solvers (denoted as ``AthK-HARM''
and ``AthK-Val'', respectively) show no major differences except for the energy accretion
rate, which is systematically lower for the Valencia solutions than in the equivalent
\HARM-like solutions. This may be related to the different treatment of the energy term
between the two formulations. However, the $64^3$ solution is clearly not in a convergent
regime, so it is not immediately clear how robust this trend is. Furthermore, though they
do not constitute the same one-to-one comparison, neither \BHAC nor \ECHO seem to be
consistently lower than \HARM-like codes at similar resolutions.

\begin{figure}
  \includegraphics[width=\columnwidth]{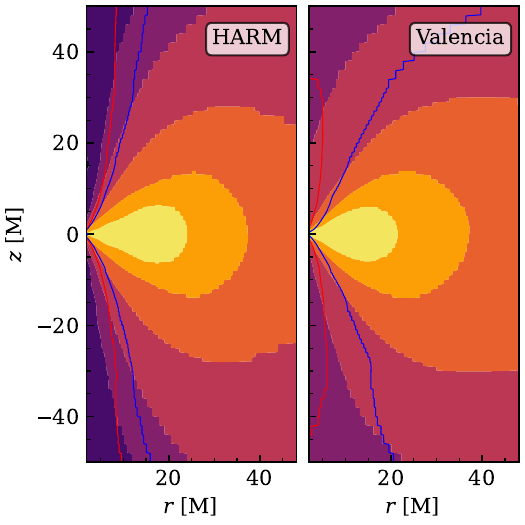}
  \caption{\label{fig:eht_az_avg_lr} Contour plots for $\rho$ azimuthally and temporally
  averaged over the range $\phi\in[0,2\pi]$, $t\in[5000M,10000M]$. The $64^3$ \HARM-like
  data is on the left, and the $64^3$ Valencia data is on the right. There are six
  logarithmically spaced bins (i.e., seven contours) in the range $[10^{-7},1]$ in units
  of $M^{-2}$. The blue contour marks $\beta^{-1}=1$, and the red contour indicates
  $\sigma=1$.}
\end{figure}
\begin{figure}
  \includegraphics[width=\columnwidth]{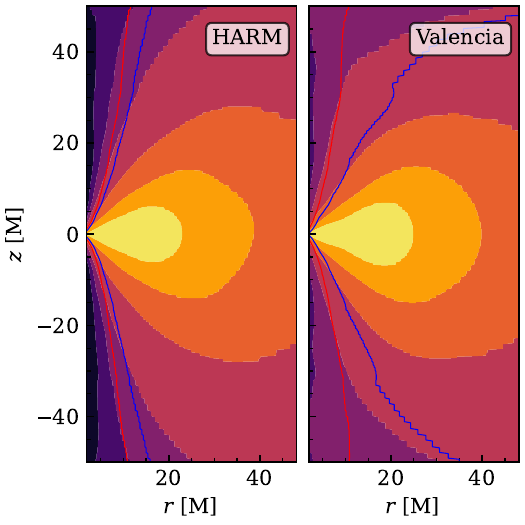}
  \caption{\label{fig:eht_az_avg_sr} Same as Figure~\ref{fig:eht_az_avg_lr} but for the
  $96^3$ data.}
\end{figure}
\begin{figure}
  \includegraphics[width=\columnwidth]{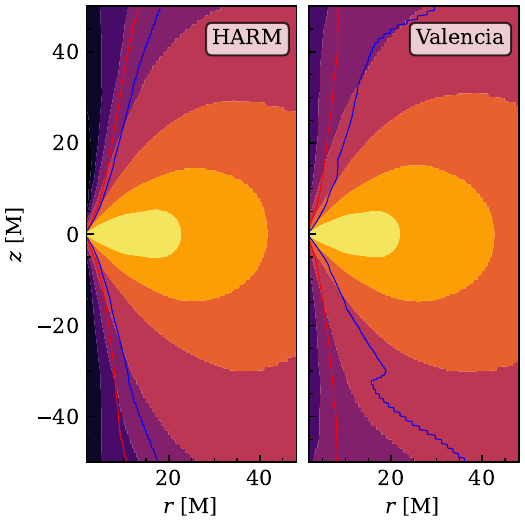}
  \caption{\label{fig:eht_az_avg_hr} Same as Figure~\ref{fig:eht_az_avg_lr} but for the
  $128^3$ data.}
\end{figure}

\begin{figure}
  \includegraphics[width=\columnwidth]{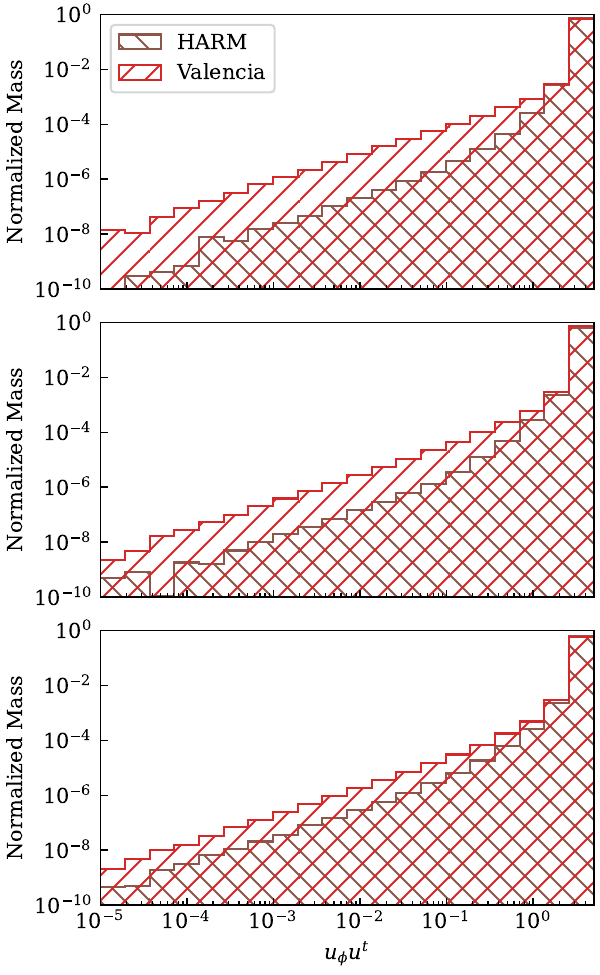}
  \caption{\label{fig:eht_lang_compare} Mass-weighted histograms of the specific angular
    momentum $u_\phi u^t$ for the $64^3$ (top), $96^3$ (middle), and $128^3$ (bottom)
    tests. Each distribution is temporally averaged over $t\in[5000M,10000M]$. The
    distributions are then normalized against the initial mass at $t=0$ and therefore do
    not add up to unity.}
\end{figure}

To better understand differences beyond the horizon,
Figures~\ref{fig:eht_az_avg_lr}--\ref{fig:eht_az_avg_hr} show contour plots of $\rho$
averaged azimuthally over $\phi\in[0,2\pi]$ and temporally over $t\in[5000M,10000M]$ for
each run. We additionally plot contours of the magnetization, $\sigma \equiv b^2/\rho$,
and the inverse plasma-beta, $\beta^{-1}\equiv b^2/2P$, at $\sigma = 1$ and
$\beta^{-1}=1$. The $\sigma$ contours loosely corresponds to the outermost boundary of the
funnel (see, e.g, \citet{McKinney:2004ka,EventHorizonTelescope:2019pcy}), while
$\beta^{-1}$ marks the equipartition region and the transition from pressure being
thermally dominated to magnetically dominated, and it often coincides with the boundary
of the disk. The density inside the funnel region decreases with improved resolution for
both solvers, with the \HARM-like solver clearly favoring lower densities compared to the
Valencia solver. We also see qualitative differences in the funnel boundary; the size of
the funnel is relatively stable in the \HARM-like formulation but steadily increases with
resolution, while the lowest contour in the Valencia formulation shows fairly significant
changes in shape and size for each resolution. We see a similar trend for $\beta^{-1}$,
where the \HARM-like formulation favors a later transition from being thermally dominated
to magnetically dominated and a much more stable shape.

The Fishbone-Moncrief torus is an equilibrium solution with uniform specific angular
momentum $\ell=u_\phi u^t$, and the only way to accrete matter or bring it into the funnel
is by material losing $\ell$. Figure~\ref{fig:eht_lang_compare}, which shows
time-averaged, mass-weighted histograms of $\ell$, suggests that the Valencia runs have
more low-$\ell$ material than the \HARM-like runs, which is in agreement with our
expectations from Figures~\ref{fig:eht_az_avg_lr}--\ref{fig:eht_az_avg_hr}. The extra
low-$\ell$ material cannot be attributed to significant differences in the amount of mass;
the $64^3$ and $128^3$ Valencia runs have ${\sim}4\%$ less mass than the \HARM-like runs,
and the $96^3$ run contains ${\sim}12\%$ more mass than the $96^3$ \HARM-like run. We also
observe that the difference in the distribution of $\ell$ decreases with resolution,
indicating that the two solvers appear to be converging toward similar results.

These differences may be linked to the lack of guaranteed energy conservation in the
Valencia formulation. If the Valencia runs slowly lose energy, the angular momentum will
also be affected, likely resulting in more low-$\ell$ material and a funnel containing
more material. This would also explain the lower energy accretion rates seen relative to
the \HARM-like formulation. However, though these tests are strongly suggestive that the
GRMHD formulation causes these differences, they are hardly conclusive. Furthermore, the
differences between either solver in \AthenaK and other GRMHD codes at similar resolutions
in the EHT data are generally larger than the differences caused by changing solvers. As
resolution increases, both solvers also appear to converge toward similar results. We
therefore conclude that one's grid setup and choice of numerical scheme are more important
than the specific GRMHD formulation.

\subsection{Binary Neutron Star Merger}
\label{subsec:bns}
Our last accuracy test duplicates the BNS merger in \citet{Cook:2023bag}, which consists
of an equal-mass binary with a total baryon mass $M_b = 3.25~\Msun$ and a gravitational
mass $M = 3.0297~\Msun$ with an initial separation of $45~\mathrm{km}$ and orbital
frequency $f_0 \approx 294~\mathrm{Hz}$. The data is constructed with the \LORENE initial
data code \citep{PhysRevD.63.064029,Gourgoulhon.8.16} assuming a polytropic EOS with
$K\approx123.6$ and $\Gamma = 2$.

The computational domain spans $\left[-1536~\Msun,1536~\Msun\right]$ in every direction
and uses seven layers of mesh refinement. We perform the test at three resolutions with
$128^3$, $192^3$, and $256^3$ cells on the base grid, which we respectively designate as
LR, SR, and HR. This corresponds to resolutions of
$\Delta x = \{0.1875~\Msun, 0.125~\Msun, 0.09375~\Msun\}$, or
$\Delta x \approx \{277~\mathrm{m}, 185~\mathrm{m}, 138~\mathrm{m}\}$, on the finest
refinement level. We ensure that refinement structure remains the same by setting the mesh
block size to $16$, $24$, and $32$ cells per dimension for the LR, SR, and HR resolutions,
respectively, resulting in $7960$ total mesh blocks per run. To facilitate comparison with
\GRAthena, we use LLF and WENOZ, and we set $\rho_\atm = 1.28\times 10^{-21}$ and
$T_\atm = 1.58255\times10^{-19}$. Unlike \GRAthena, \AthenaK does not currently have an
apparent horizon finder which can be used for excising the fluid inside a black hole. In
order to ensure stability through collapse, we use an ad-hoc excision scheme by flooring
all fluid variables inside the region $\alpha < 0.20$. Also differently from \GRAthena, we
enable FOFC and set $f_\thr=0.1$ Neither of these choices will strongly affect the
waveform, but they improve mass conservation and the overall stability of our solution.
However, we do not enable the DMP scheme, as this particular binary system should be
well-behaved even with standard numerical methods. We also note that the \GRAthena run
contains a poloidal magnetic field, which the \AthenaK runs neglect. However, the magnetic
field is not dynamically important for gravitational waves during the inspiral and early
post-merger even for very large fields
\citep{Ioka:2000yb,Giacomazzo:2009mp,Palenzuela:2015dqa, Palenzuela:2022kqk}, so this
difference should not affect our comparison.

As an additional comparison, we also perform an additional run with \WhiskyTHC
\citep{Radice:2012cu, Radice:2013hxh, Radice:2013xpa, Radice:2015nva} using a high-order
conservative finite-difference method with MP5 reconstruction. \WhiskyTHC is based on the
\EinsteinToolkit \citep{Loffler:2011ay, EinsteinToolkit:2024_05} and on the \code{Carpet}
AMR driver, which implements Berger-Oliger patch-based refinement \citep{Berger:1984zza}.
Note that refluxing is not supported for the finite-differencing version of \WhiskyTHC
used here.  We evolve the spacetime using the \code{CTGamma} code \citep{Pollney:2009yz,
Reisswig:2013sqa}, which solves the Z4c formulation of Einstein's equations, like
\AthenaK. The \WhiskyTHC simulations have 7 refinement levels. The finest one covers each
star during inspiral and has a grid spacing $h = 0.125\ M_\odot$. Time integration is
performed with a 3rd-order strongly-stability preserving Runge-Kutta scheme. The CFL for
the \WhiskyTHC simulations is set to $0.125$, to be able to use the positivity-preserving
limiter in \WhiskyTHC \citep{Radice:2013xpa}. 

\begin{figure}
  \includegraphics[width=\columnwidth]{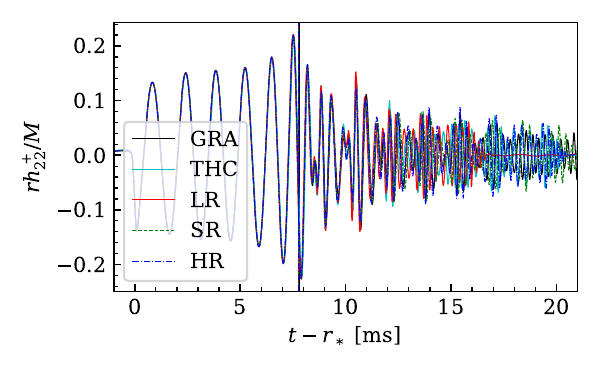}
  \includegraphics[width=\columnwidth]{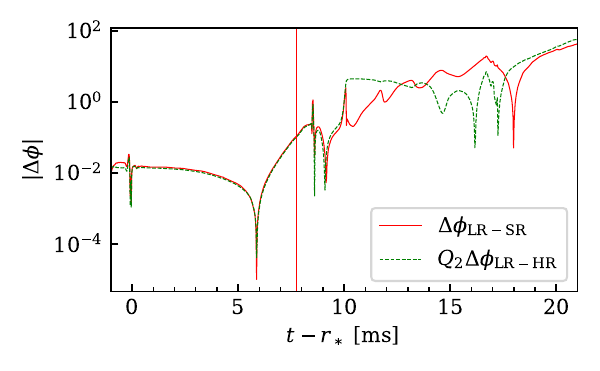}
  \caption{\label{fig:gw_strain} (Top) The real part of the gravitational wave strain in
  the $(2,2)$ mode, normalized by extraction radius and gravitational mass. We compare our
  LR, SR, and HR \AthenaK runs to a comparable run in \GRAthena. Vertical lines show
  merger time as estimated by the peak amplitude of $h^+_{22}$. (Bottom) Phase convergence
  of the three \AthenaK runs. The difference between $\phi_\mathrm{LR}$ and
  $\phi_\mathrm{HR}$ is rescaled by a factor $Q_2$ assuming second-order convergence.}
\end{figure}

We show the gravitational waveform extracted at $R=500~\Msun$ (except for the \WhiskyTHC
run, which was extracted at $R=400~\Msun$) in the top half of Figure~\ref{fig:gw_strain}
as a function of the retarded time, $t-r_\ast$, where $r_\ast=r+2M\log(r/2M - 1)$ and 
$r(R) = (1 + M/2R)^2 R$. Our LR solution corresponds to the $128^3$ solution for \GRAthena
reported in \citet{Cook:2023bag}. We find excellent agreement between these two solutions
through merger, and agreement still remains reasonable through the first several
milliseconds of the post-merger phase. The difference in merger time is $\Delta t \approx
0.05$~ms, with \AthenaK merging slightly earlier. The SR solution roughly corresponds to
the \WhiskyTHC results, and the two are in good agreement despite using different
numerical methods; the difference in merger time between the two is
$\Delta t \approx 0.01$~ms and sits approximately halfway between the LR run and
\GRAthena.

We do not see good agreement in collapse time; \AthenaK's LR solution collapses
${\sim}9$~ms after merger, while the \GRAthena run does not collapse in the plotted
duration. Similarly, \WhiskyTHC predicts collapse around ${\sim}13$~ms after merger, but
the SR solution does not collapse during this time. There is also no consistency in the
\AthenaK runs in this region, with the HR solution collapsing after the LR run but before
the SR case. However, collapse time is notoriously sensitive to perturbations, and even
very similar numerical methods predict different collapse times (e.g.,
\citet{Espino:2022mtb}).

We use the SR and HR resolution runs to establish convergence of the waveform phase. To
eliminate phase offsets caused by non-converging noise prior to the arrival of the
so-called ``junk'' radiation at the extraction radius, we align the waveforms using the
procedure in \citet{Boyle:2008ge} between $t-r_\ast=0$ and the merger time of the LR
waveform. We show the results in the bottom plot in Figure~\ref{fig:gw_strain}. Throughout
the inspiral and merger, we maintain approximate second-order convergence as expected by
our numerical methods. The results do not show clean convergence in the post-merger phase
even before collapse. However, we note that most NR codes struggle to maintain convergence
in the post-merger phase, especially with second-order schemes \citep{Bernuzzi:2011aq,
Radice:2015nva,Most:2019kfe,Zappa:2022rpd}. Though going to higher resolution may improve
convergence, achieving second-order convergence is unlikely due to the presence of fluid
turbulence, which, without including an explicit viscous term, will cascade to smaller
length scales as resolution is increased.

\begin{figure}
  \includegraphics[width=\columnwidth]{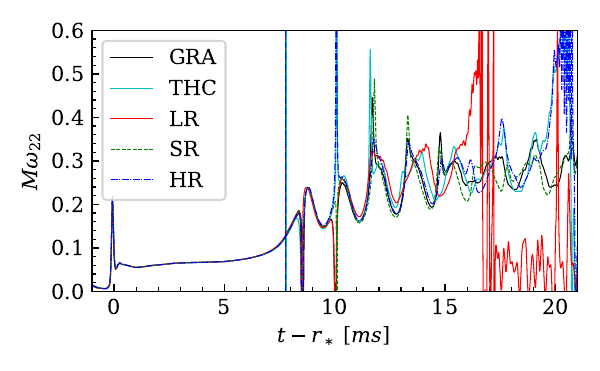}
  \includegraphics[width=\columnwidth]{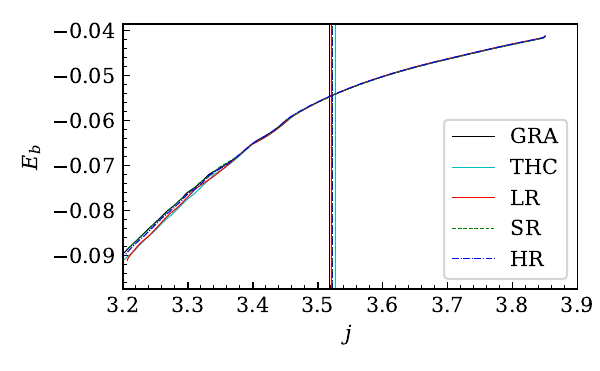}
  \caption{\label{fig:gw_energetics} (Top) Instantaneous frequency of the $(2,2)$ mode,
  $\omega_{22}$, for \GRAthena and the \AthenaK runs. (Bottom) A plot of the reduced
  binding energy $E_b$ and reduced angular momentum $j$. In both plots, vertical lines
  indicate merger times for each model.}
\end{figure}

As suggested by Figure~\ref{fig:gw_strain}, Figure~\ref{fig:gw_energetics} confirms that
\AthenaK agrees very well in the instantaneous frequency, $\omega_{22}=\dot{\phi}_{22}$,
with both \GRAthena and \WhiskyTHC through merger. The reported frequencies at merger
are $\omega_{22}^\mathrm{GRA}\approx1389~\Hz$, $\omega_{22}^\mathrm{THC}\approx1329~\Hz$,
$\omega_{22}^\mathrm{LR}\approx1377~\Hz$, $\omega_{22}^\mathrm{SR}\approx1359~\Hz$, and
$\omega_{22}^\mathrm{HR}\approx1353~\Hz$, which are all consistent with each other.

We next consider the reduced binding energy,
$E_b = (M_\mathrm{ADM} - E_\mathrm{GW} - M)/(M\nu)$, as a function of the reduced angular
momentum, $j = (J_\mathrm{ADM} - J_\mathrm{GW})/(M^2 \nu)$, where for our initial data
$M_\mathrm{ADM} = 2.9984~\Msun$, $J_\mathrm{ADM} = 8.83542~\Msun^2$ in natural units, and
the symmetric mass ratio is $\nu = q/(1 + q)^2 = 1/4$. This is a gauge-invariant measure
which shows the evolution of the binary as it loses both energy and angular momentum due
to gravitational radiation \citep{Damour:2011fu,Bernuzzi:2012ci}. The bottom plot in
Figure~\ref{fig:gw_energetics} shows that our runs are all in excellent agreement prior to
merger. The lines do diverge somewhat in the post-merger phase, but the differences in the
\AthenaK runs decrease with resolution, indicating some measure of convergence.

\begin{figure}
  \includegraphics[width=\columnwidth]{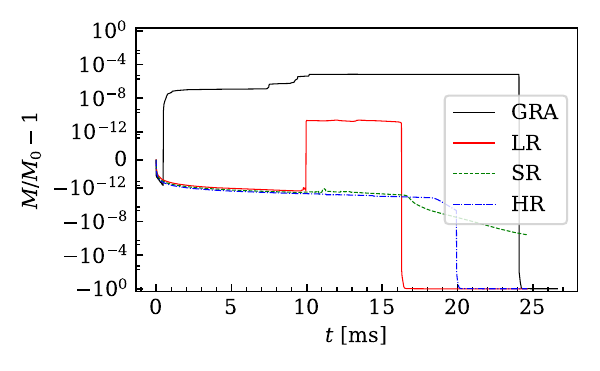}
  \caption{\label{fig:bns_mass} Mass conservation of \AthenaK BNS runs compared to
  \GRAthena.}
\end{figure}

Due to the effects of round-off error and the artificial atmosphere, mass will not be
perfectly conserved in GRMHD evolutions. Figure~\ref{fig:bns_mass} shows the mass
conservation of BNS runs in \AthenaK compared to \GRAthena. Prior to merger, \AthenaK
demonstrates violations in mass conservation at ${\sim}10^{-11}$ or better. Though the
sign of this error changes, the magnitude of conservation violation remains at roughly
this same order after merger until collapse for the LR case. This represents an
improvement in conservation of $\mathcal{O}(10^4)$ or better over \GRAthena. The SR and
HR cases do begin to demonstrate large mass losses several milliseconds after merger.
These errors, however, are simply due to matter outflows passing the edge of the
computational domain.

Mass violations prior to merger seem to scale linearly with $t$, suggesting there is a
small but consistent mass loss. This cannot be blamed on primitive inversion failures, as
there are exactly two failures across all three runs (both of which occurred early in the
HR run at points very close to atmosphere). We also cannot blame this on our flooring
policy, which almost always results in an \textit{increase} in mass. Though floating-point
errors typically scale as $\sqrt{t}$, this is based on the assumption that floating-point
errors are equally likely to round up as they are to round down. If such errors were
instead consistently biased downward, the behavior would be approximately linear instead.
One possibility leading to linear behavior may be truncation errors caused by mass fluxes
between the atmosphere and the stellar surface or its outflows. However, we do not have
sufficient evidence to confirm that this is the case.

\section{Scaling and Performance}
\label{sec:scaling}
To evaluate the performance of \AthenaK at large scales, we perform weak and strong
scaling tests on OLCF Frontier, which has four AMD MI250X per node (each with two compute
dies, or eight logical GPUs per node). Because scaling is highly problem-dependent, we
provide two different test problems.

The first test problem tests only the Valencia GRMHD solver on a uniform grid in a fixed
spacetime in a setup designed to explore the growth of turbulence, particularly the
magnetorotational instability. The initial data is a Fishbone-Moncrief torus around a
Kerr black hole with dimensionless spin $a=0.9375$, similar to our setup used in
Section~\ref{subsec:fm_disk}. We zoom into a small cube located of width
${\sim}\left(M/\Msun\right)~\mathrm{km}$ centered on $x=9M$, $y=z=0$ and fix the spacetime
to the appropriate Kerr-Schild data. The strong scaling test has $8192$ cells per
dimension for a resolution of $\Delta x \approx \left(M/\Msun\right) 11.9~\mathrm{cm}$,
which requires a minimum of 1024 nodes. To perform weak scaling tests starting at a single
node, we reduce the resolution to $1024\times1024\times512$ cells.

The second test problem consists of a single unperturbed TOV star in a dynamical spacetime
similar to Section~\ref{subsec:free_ns}. To better emulate a typical BNS postmerger setup,
we place it on an extended grid spanning $[-1024~\Msun,1024~\Msun]$ in all directions with
a base resolution of $256^3$ points and six levels of refinement, with each successively
halving the size of the domain. Therefore, the finest level has a resolution of
$0.125~\Msun$ (${\sim}185~\mathrm{m}$) on a region spanning $[-16~\Msun,16~\Msun]$ in all
directions, comparable in resolution to the SR run in \ref{subsec:bns}.

For all tests, we use RK3, HLLE, and WENOZ, and we additionally enable FOFC for the
turbulence setup but leave it disabled in the TOV/BNS setup (which incurs roughly a $3\%$
to $5\%$ performance decrease in our TOV tests above). We also disable all file output and
diagnostic calculations, including wave extraction. Each test runs for 20 time steps, and
performance is measured in terms of zone-cycles (or cell updates) per second.

\subsection{Weak Scaling}
\label{subsec:weak}

\begin{figure}
  \includegraphics[width=\columnwidth]{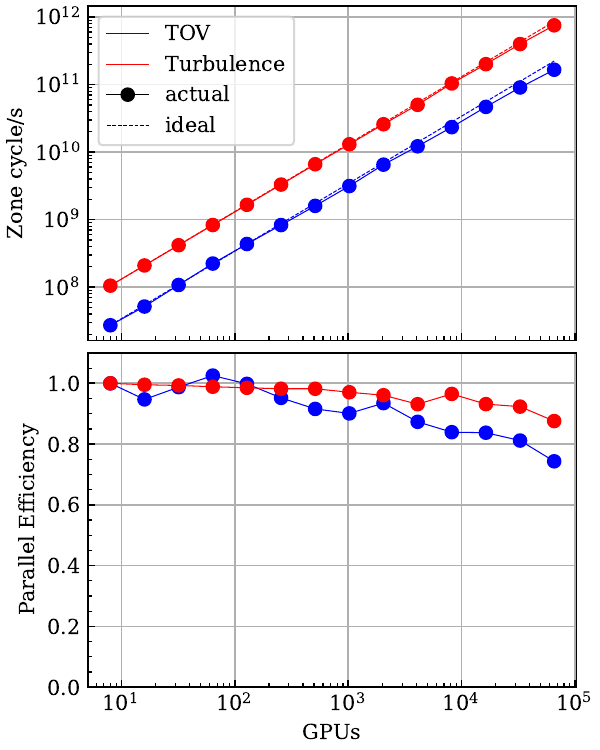}
  \caption{\label{fig:weak_scaling} Weak scaling results of a single neutron star (blue)
  and a turbulence setup (red) on OLCF Frontier. The top plot displays raw performance
  (solid lines) in terms of zone-cycles per second relative to ideal scaling (dashed
  lines). The bottom plot shows parallel efficiency relative to a single node as a
  function of GPU count.}
\end{figure}

In a weak scaling test, the problem size per node is kept fixed, i.e., resolution is
increased to match computing resources. This measures the ability of a code to perform
large-scale calculations before parallel communication becomes excessive. For each
iteration of the test, we double the resolution in one dimension as the number of nodes is
doubled. The turbulence test uses mesh blocks with $256^3$ cells to maximize performance,
but the TOV uses smaller mesh blocks of $64^3$ cells to maintain the appropriate grid
structure. We show results up to 65,536 GPUs (8192 nodes) in
Figure~\ref{fig:weak_scaling}. The turbulence setup maintains $\geq 87\%$ efficiency up
to 65,536 GPUs, and the TOV setup maintains $\geq 81\%$ efficiency up to 32,768 GPUs and
$\geq 74\%$ efficiency up to 65,536 GPUs.

The turbulence setup is ${\sim}4$ times faster than the TOV setup. Profiling the code
suggests that the Z4c solver is slightly slower than the GRMHD solver, which accounts for
more than half of the slowdown. The mesh refinement, particularly the high-order
prolongation and restriction operators, incur an additional performance penalty.

The remaining performance difference is related to the size of the mesh blocks. Smaller
mesh blocks offer better flexibility for block-based SMR and AMR, but larger mesh blocks
generally have improved performance. This occurs for two reasons: firstly, fewer mesh
blocks means less interblock communication as there are fewer ghost cells relative to the
number of physical cells. Secondly (and more importantly), the innermost
level of vectorization in the GRMHD solver operates on single lines in the $x$ direction.
This is convenient for reconstruction operators, which are most easily written to
reconstruct one side of two different interfaces, and it generally offers better CPU
performance. However, GPU performance becomes very closely tied to the length of the mesh
block in the $x$ direction as a result. Consequently, the larger mesh blocks in the
turbulence test offer a modest performance boost over the smaller blocks used in the TOV
test.

\subsection{Strong Scaling}
\label{subsec:strong}

\begin{figure}
  \includegraphics[width=\columnwidth]{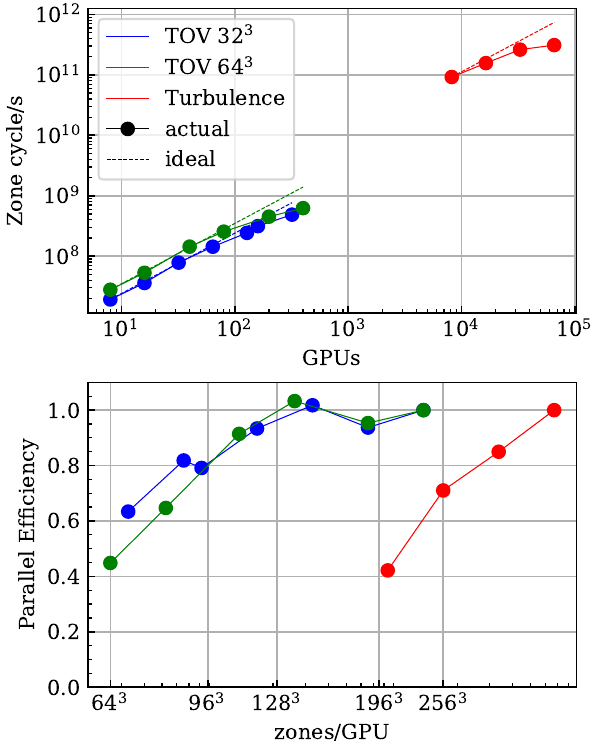}
  \caption{\label{fig:strong_scaling} The same as Figure~\ref{fig:weak_scaling}, but for
  strong scaling instead, and with efficiency measured against zones per GPU rather
  than total GPU count.}
\end{figure}

Strong scaling measures how increasing computational resources speeds up the calculation
of a fixed problem size. Though strong scaling does demonstrate the effects of increased
communication overhead, it also measures how large of a problem is needed to achieve
optimal performance. A GPU requires relatively heavy workloads both to saturate the
device and to negate the overhead of launching a new kernel. However, GPUs generally
contain far less memory per TFLOP than CPU-based machines; this makes strong scaling tests
more difficult on GPUs, as there is less difference between a saturating workload and the
maximum possible workload.

For the TOV test, we perform tests with mesh block sizes of $32^3$ and $64^3$. As
mentioned above, the mesh block size is closely related to the performance of the GRMHD
solver, so these two tests provide a way to quantify this effect in configurations
suitable for a BNS remnant. The turbulence test is done with a single configuration using
mesh blocks of $128^3$ each, which offers marginally worse performance ($<1\%$ decrease)
than the larger mesh blocks used in Section~\ref{subsec:weak}, but it allows us to scale
to larger node counts.

We show the strong-scaling test results in Figure~\ref{fig:strong_scaling}. Relative to a
single node (${\sim}236^3$ zones per GPU), \AthenaK on Frontier maintains $\geq 80\%$
efficiency down to ${\sim}96^3$ zones per GPU for the TOV test. The $64^3$ block test is
roughly $45\%$ faster than the $32^3$ test, suggesting that $32^3$ mesh blocks do not
saturate the GPUs properly. Despite the performance difference, scaling results are
comparable between the two setups at smaller node counts. However, there is some evidence
suggesting that large mesh blocks scale more poorly at large node counts. Though the
limited tests shown here are hardly conclusive, this seems reasonable; larger mesh blocks
improve GPU vectorization and reduce the total number of communication calls, but the cost
of launching a kernel does not substantially change, meaning that kernel overhead will
become a bottleneck sooner than in the case of small mesh blocks.

Strong scaling for the turbulence test is not as good, with efficiency decreasing to
$71\%$ after only quadrupling the number of resources (going from ${\sim}406^3$ to
${\sim}256^3$ zones per GPU). This is not unexpected, as this test lacks both the Z4c
solver and the computational overhead associated with mesh refinement.

\subsection{Discussion on Performance}
\label{subsec:performance}
Our scaling results are best compared to \Parthenon, a similar \Kokkos-based AMR framework
derived from \AthenaPP \citep{Grete.5.2023}. Our turbulence weak scaling results are
similar to those reported for \Parthenon on Frontier. We cannot draw a direct comparison
for the TOV tests, as \Parthenon does not have weak scaling results for its mesh
refinement. Strong scaling is noticeably poorer in all cases considered, but a likely
explanation for this is the additional memory footprint required by our tests limiting the
size of the tests we can run; compared to standard hydrodynamics, which need only store
the conserved and primitive variables, MHD requires additional memory for both the face
and cell-centered magnetic fields. The GRMHD implementation presented here also stores the
ADM variables to support generic spacetimes, even when such spacetimes are unevolved.
Therefore, they consume significant memory but do not contribute to the overall
computational load.

\begin{table}
  \begin{ruledtabular}
    \caption{\label{tab:perf} Summary of AthenaK performance for a freely evolving TOV
      on different architectures. Speedup is measured relative to a single 64-core CPU
      socket on an AMD EPYC 7763. For reference, \GRAthena results on this system are
      also included.}
    \begin{tabular}{l|cc}
      Architecture & $10^6$~zone-cycles/s & Speedup \\
      \hline
      EPYC 7763 & $0.760$ & $1$ \\
      EPYC 7763 (\GRAthena) & $0.964$ & $1.27$ \\
      A100 & $10.7$ & $14.1$ \\
      MI250X & $4.84$ & $6.37$ \\
      Ryzen 9 5900X & $0.125$ & $0.164$ \\
      RTX 3070 (laptop) & $1.08$ & $1.42$
    \end{tabular}
  \end{ruledtabular}
\end{table}

In addition to the scaling results above, we performed a small number of tests comparing
CPU and GPU performance on NERSC Perlmutter, which consists of CPU nodes (each with two
AMD EPYC 7763 CPUs) and GPU nodes (each with a single CPU and four Nvidia A100 GPUs). Our
setup is similar to Sec.~\ref{subsec:free_ns} but with all output disabled and limited to
the first 20 time steps. We also reduce the resolution to $128^3$ with mesh blocks of size
$32^3$ to ensure they fit on a single 40~GB GPU. Running with a single CPU (64 cores/128
threads) achieves $7.60\times10^5$~zone-cycles/s, and a single Nvidia A100 GPU manages to
achieve $1.07\times10^7$~zone-cycles/s, a factor of ${\sim}14$ improvement. The same
setup on OLCF Frontier using a single MI250X compute die achieves
$4.84~\times10^6$~zone-cycles/s. To provide a baseline comparison, we also perform this
test with \GRAthena using as similar a setup as possible. For a cell-centered spacetime,
\GRAthena achieves $9.64\times10^5$~zone-cycles/s, which is somewhat faster but comparable
to \AthenaK's CPU performance.

We also ran a version of this test on a consumer-grade laptop with an Nvidia RTX 3070
laptop GPU. Due to the reduced amount of memory, we reduced the resolution by half,
switched to $16^3$ mesh blocks, and imposed a reflection symmetry across the z-axis, but
the physics options are the same. The laptop GPU achieves $1.08\times10^6$~zone-cycles/s,
which is $42\%$ faster than a single CPU on Perlmutter. The same calculation using the
laptop's CPU (an AMD Ryzen 9 5900X) with 16 OpenMP threads only achieved
$1.25\times10^5$~zone-cycles/s. All of these results are summarized in
Table~\ref{tab:perf}. We also found it possible to run a very coarse
(${\sim}900~\mathrm{m}$) BNS simulation through merger on this same device in a matter of
a few hours. Though research-grade calculations still require larger machines, the
order-of-magnitude speedup provided by GPUs makes it possible to accelerate testing and
debugging considerably.


\section{Conclusion}
\label{sec:conclusion}
In this paper, we have introduced an extension to \AthenaK's GRMHD capabilities to
dynamical spacetimes. Our solver uses the Valencia formulation and finite-volume methods,
and we incorporate an FOFC scheme with optional DMP enforcement to improve our atmosphere
treatment enhance overall stability.

We have performed a number of standard tests in both flat and curved
spacetimes. \AthenaK is in excellent agreement with analytical results where they exist
and compares favorably with other GRMHD codes. In particular, our cylindrical blast wave
test demonstrates the ability of the FOFC scheme to improve robustness when handling
strongly magnetized flows, which enables us to evolve a strongly magnetized blast wave
with WENOZ. We also accurately reproduce the oscillation frequencies of a TOV star in both
the Cowling approximation and full GR, and we show that the FOFC scheme does not adversely
affect this evolution.

To validate \AthenaK's magnetic field evolution in curved spacetimes, we compare the
results of a SANE accretion disk around a Kerr black hole to several of the GRMHD codes in
the EHT code comparison project \citep{EventHorizonTelescope:2019pcy}. \AthenaK achieves
similar results, validating the GRMHD solver on a highly nontrivial problem. This test
also allows a consistent comparison between \AthenaK's standard \HARM-like GRMHD solver
and the new solver we have added for dynamical spacetimes based on the Valencia
formulation. We find some small qualitative differences which may be due to their
different energy terms. Nevertheless, these differences seem to be minor, and the
comparison with other GRMHD codes suggests that one's choice of numerical methods and grid
structure have a far larger effect on the evolution.

We also perform unmagnetized BNS simulations on GPUs, which reproduce the results shown in
\citep{Cook:2023bag} with \GRAthena and are in excellent agreement with \WhiskyTHC. A
resolution study shows that \AthenaK's gravitational waves achieve second-order phase
convergence, and we are able to evolve stably through collapse. Thanks to the FOFC scheme,
we also show that \AthenaK improves mass conservation over \GRAthena by a factor of
$\mathcal{O}(10^4)$.

The performance of \AthenaK is also very good, with GPUs demonstrating an
order-of-magnitude speedup over CPUs. We also only demonstrate a small slowdown on CPUs
relative to a cell-centered version of \GRAthena despite optimizing our vectorization for
GPU performance. We maintain $74\%$ weak scaling efficiency up to 65,536 GPUs and better
than $80\%$ up to 32,768 GPUs on OLCF Frontier for a dynamical GRMHD problem. We also
maintain better than $70\%$ strong scaling efficiency down to ${\sim}96^3$ cells per GPU,
which corresponds to less than 4 $64^3$ mesh blocks or 27 $32^3$ mesh blocks per GPU. Our
fixed-spacetime turbulence tests achieve better than $87\%$ efficiency weak efficiency up
to 65,536 GPUs.

We are also working to expand \AthenaK's physics capabilities. Microphysical tabulated
equations of state are being tested, and we are also developing a new neutrino transport
module based on a finite-element scheme \citep{Bhattacharyya:2022bzf}. Nevertheless, even
without these enhancements, \AthenaK's exascale-ready performance and excellent stability
properties make it a capable tool for NR applications such as precision waveform modeling.

\begin{acknowledgments}
The authors thank the Penn State Numerical Relativity Group for helpful discussions and
important feedback. This research makes use of data from the EHT Code Comparison Project,
and JF thanks Oliver Porth for assistance with accessing the data.

Part of this work was completed at the TACC Open Hackathon, part of the Open Hackathons
program. The authors would like to acknowledge OpenACC-Standard.org for their support. We
are particularly grateful to Matthew Cawood, Victor Eijkhout, and Forrest Glines for
helpful discussions while optimizing \AthenaK.

SB and BD acknowledge support by the EU Horizon under ERC Consolidator Grant, no. InspiReM-101043372.

This research was supported by funding from the U.S. Department of Energy, Office of
Science, Division of Nuclear Physics under Award Number(s) DE-SC0021177 and DE-SC0024388,
by NASA under award No. 80NSSC21K1720, and by the National Science Foundation under Grants
No. PHY-2011725, PHY-2020275, PHY-2116686, and AST-2108467.

Part of this work was completed during a sabbatical visit to the Institute for Gravitation
and the Cosmos at the Pennsylvania State University by SB.

Simulations were performed on OLCF's Frontier, NERSC's Perlmutter, and on the Pennsylvania
State University's Institute for Computational and Data Sciences's Roar Collab
supercomputer.  This research used resources of the National Energy Research Scientific
Computing Center, a DOE Office of Science User Facility supported by the Office of Science
of the U.S.~Department of Energy under Contract No.~DE-AC02-05CH11231.  This research used
resources of the Oak Ridge Leadership Computing Facility at the Oak Ridge National
Laboratory, which is supported by the Office of Science of the U.S. Department of Energy
under Contract No. DE-AC05-00OR22725.

\end{acknowledgments}

\bibliography{references}

\end{document}